\RequirePackage{fix-cm}
\documentclass[smallextended]{svjour3}       
\smartqed  
\usepackage{graphicx}
\usepackage{natbib}
\usepackage{siunitx}
\usepackage[version=4]{mhchem}
\usepackage{lscape}

\newcommand\micron{$\mu$m}

\usepackage{soul}
\usepackage{color}

\usepackage{pdfcomment}
\usepackage{hyperref}
\usepackage{ulem}

\DeclareSIUnit\mols{molec.\,s^{-1}}
\newcommand{\herschel}{Herschel}
\newcommand{\trans}{$1_{10}\text{--}1_{01}$}

\providecommand\aap{A\&A}%
\providecommand\apjl{ApJ}%
\providecommand\apjs{ApJS}%
\begin{document}

\title{The Main Belt Comets and Ice in the Solar System}

\author{Colin Snodgrass \and
Jessica Agarwal \and
Michael Combi \and
Alan Fitzsimmons \and
Aurelie Guilbert-Lepoutre \and
Henry H.~Hsieh \and
Man-To Hui \and
Emmanuel Jehin \and
Michael S.~P.~Kelley \and
Matthew M.~Knight \and
Cyrielle Opitom \and
Roberto Orosei \and
Miguel de~Val-Borro \and
Bin Yang
}

\institute{C.~Snodgrass \at
              School of Physical Sciences, The Open University, Milton Keynes, MK7 6AA, UK\\
              Tel.: +44-1908-654320\\
              \email{colin.snodgrass@open.ac.uk}           
           \and
              J.~Agarwal \at Max-Planck-Institut f{\"u}r Sonnensystemforschung (Germany) \and
              M.~Combi \at University of Michigan (USA) \and
              A.~Fitzsimmons \at Queen's University Belfast (UK) \and 
              A.~Guilbert-Lepoutre \at CNRS/UTINAM - UMR 6213 UBFC, Besan\c{c}on (France) \and 
              H.~H.~Hsieh \at Planetary Science Institute (USA); Academia Sinica (Taiwan) \and 
              M.-T.~Hui \at University of California Los Angeles (USA) \and 
              E.~Jehin \at Universite de Liege (Belgium) \and 
              M.~S.~P. Kelley \& M.~M.~Knight \at University of Maryland (USA) \and 
              C.~Opitom \& B.~Yang \at European Southern Observatory (Chile) \and 
              R.~Orosei \at Istituto Nazionale di Astrofisica (Italy) \and 
              M.~de~Val-Borro \at NASA Goddard Space Flight Center (USA)
}

\date{Received: date / Accepted: date}

\maketitle

\begin{abstract}
We review the evidence for buried ice in the asteroid belt; specifically the questions around the so-called Main Belt Comets (MBCs). We summarise the evidence for water throughout the Solar System, and describe the various methods for detecting it, including remote sensing from ultraviolet to radio wavelengths. We review progress in the first decade of study of MBCs, including observations, modelling of ice survival, and discussion on their origins. We then look at which methods will likely be most effective for further progress, including the key challenge of direct detection of (escaping) water in these bodies. 
\end{abstract}

\section{Introduction}

The traditional view of our Solar System neatly divides it into the inner part, home of the terrestrial planets and rocky asteroids, and the outer region of the gas giants and icy small bodies. These are separated by the `snow line', which marks the distance from the Sun where the ambient temperature allows icy bodies to form and survive. In this picture, the terrestrial planets formed `dry', as only rocky material condensed from the solar nebula in the inner regions, while the outer planets became giants due to the fast formation of icy cores, followed by the run-away accretion of the abundant gas further from the young Sun. Earth's water was then delivered by occasional impacts of comets, whose eccentric orbits brought ice from their distant parent regions to the terrestrial planet region.

Despite the pleasing simplicity of this model, there are a number of awkward problems: The samples of asteroids that we have on Earth, meteorites, show a variety of compositions, including aqueously altered minerals which must have formed in the presence of water, incompatible with asteroids being entirely dry rocks \citep[e.g.,][]{Brearley+Jones1998}. On the other hand, results from the Stardust mission showed that comet dust contains minerals formed at high temperature, presumably near to the Sun \citep{Brownlee2006}. Meanwhile, many extrasolar systems with gas giant planets close to their stars have been discovered, presenting a challenge to formation models \citep[e.g.,][]{Mayor+Queloz,2015ARA&A..53..409W}. Finally, recent observations have uncovered evidence of ice in unexpected places in the inner Solar System, including the population of `Main-Belt Comets' (MBCs) which have stable asteroid-like orbits, inside the snow line, but which demonstrate comet-like activity \citep{hsieh06}.

The solution comes from the recognition that planetary systems are dynamic places, with the orbits of even the largest planets able to evolve and migrate, especially in the early period when interaction between forming planets and the proto-planetary disc is strong. Numerical models that trace how planetary orbits interact and change can explain the migration of planets to Hot Jupiter orbits \citep[e.g.,][]{trilling1998}. More importantly, for studies of our own Solar System, these models also reveal the way that such planetary migrations stir up the population of small bodies. They can therefore be tested, by comparing their outcomes with the observed architecture of the Solar System -- not only the present-day orbits of the planets, but also the orbits and different properties of various populations of comets and asteroids. 
For example, the `Nice model' \citep{gomes05} is able to reconstruct the architecture of the trans-Neptunian region, possibly implant trans-Neptunian objects (TNOs) in the main asteroid belt \citep{levison09}, and explain the increased impact rate on the Moon during the Late Heavy Bombardment period, via the gravitational effect of Uranus and Neptune interacting. The latest family of models, known as the `Grand Tack'  \citep{walsh11}, uses Jupiter and Saturn migrating first inwards and then `tacking' outwards to explain the relatively small size of Mars and also to scatter both rocky and icy small bodies throughout the Solar System.

A common criticism of these dynamical models is that they are highly tunable and therefore lack predictive power -- the input parameters can be arbitrarily adjusted until they produce a simulated Solar System that looks like our own. Increasing the number of independent constraints on these models, thus reducing the amount of available free parameter space, is therefore an important way to refine and improve them.  For example, a recent analysis of meteoritic evidence \citep{doyle2015} appears to rule out the possibility that the parent bodies of carbonaceous chondrite meteorites (i.e., C-type asteroids) were formed beyond Jupiter, as suggested in the Grand Tack model, indicating that at least this feature of the model, if not the whole model itself, is inconsistent with the physical evidence.  Study of the present-day distribution of icy bodies, particularly those containing water ice, is another way to provide such constraints on these models, which generally predict that icy bodies will be found everywhere, but differ in details such as abundance, distribution, and ratios relative to other materials (e.g., rocky components or other volatiles).

In this review, we look at the MBCs as a population of icy bodies. We consider them in the context of water and ice detections throughout the Solar System (Section~\ref{sec:ice-everywhere}), reviewing what is known about this population in general (Section~\ref{sec:MBCs}) before considering the specific problems of modeling ice survival in their interiors (Section~\ref{sec:ice-survival}) and of directly studying this ice observationally (Section~\ref{sec:water-obs}), including predictions for activity levels (Section~\ref{sec:MBC-expected-levels}). We also consider what the lessons learned about comets in general from Rosetta tell us about MBCs (Section~\ref{sec:Rosetta}) before discussing what future observations and missions will further advance this field (Section~\ref{sec:future}). 

Previous reviews on the subject of MBCs cover the more general topic of `active asteroids', including considering how non-icy bodies can eject dust and therefore exhibit comet-like appearances \citep{Bertini2011,jewitt15}. We briefly discuss this below, but concentrate on questions related to MBCs and ice in the Solar System. The topic of water in small bodies, and in the Solar System more generally, is also the subject of earlier reviews by \citet{Jewitt2007} and \citet{encrenaz2008},  while  \citet{Hartmann2017}  consider the topic of water in extrasolar protoplanetary discs. \citet{Dones2015} give a recent review of the various cometary reservoirs in our Solar System from a dynamical point of view.


\section{Ice in the Solar System}\label{sec:ice-everywhere}

Water, usually in the form of ice, is found throughout the Solar System. Beyond Earth, it has long been recognised in the outer planets and comets, and is now also observed throughout the terrestrial planet region. Evidence for water is found with now almost monotonous regularity on Mars, but more surprisingly, ice has also been identified in permanently shadowed craters of Mercury \citep{Lawrence2013} and the Moon \citep{Colaprete2010}. While these deposits could plausibly have been delivered by comet impacts in the geologically recent past, evidence for ice in smaller bodies is more difficult to attribute to an exogenous source. 

\subsection{Water in the planets}

Radar mapping of Mercury suggested the presence of polar ice in 1991 \citep{slade-butler1992,harmon-slade1992}. Thermal models show that in permanently shadowed regions of high-latitude craters, water ice covered by a regolith layer can be stable to evaporation over billions of years \citep{paige-wood1992,vasavada-paige1999}. The ice is thought to have been implanted by either constant micrometeoritic, asteroidal and cometary influx \citep{killen-benkhoff1997}, or to stem from a few large impacts by comets and/or asteroids \citep{moses-rawlins1999,barlow-allen1999}. The MESSENGER spacecraft observed areas of high and low near-infrared (NIR) reflectivity, which is interpreted as surface ice and ice buried under a layer of organic material \citep{neumann-cavanaugh2013,paige-siegler2013}. The total amount of polar water ice on Mercury is estimated to 3$\times$10$^{15}$\,kg \citep{eke-lawrence2017}, equivalent to 300 comets the size of 67P/Churyumov-Gerasimenko\footnote{Hereafter 67P. We will give the name of comets only when they are first mentioned.} \citep{Paetzold2016}.

Water ice has also been hypothesised to exist on the Moon in permanently shadowed craters near the poles \citep{spudis-bussey2013,hayne-hendrix2015}, although a debate about alternative interpretations of data is on-going \citep{eke-bartram2014,haruyama-yamamoto2013}. Hydroxyl- and/or water-bearing materials are widely spread across the lunar surface \citep{pieters:2009}.

On Venus, water has been found only in the form of atmospheric vapour in spurious quantities of the order of a few parts per million in the nitrogen- and CO$_2$-dominated atmosphere \citep{encrenaz2008}. The high deuterium to hydrogen (D/H) ratio \citep{debergh-bezard1991} in the Venusian atmosphere is interpreted as an indication for an earlier escape of water to space from the upper atmosphere, which would be more efficient for the lighter isotope. The absence of water from the Venusian atmosphere has been connected to the strong greenhouse effect that makes the existence of ice or liquid water on Venus unlikely \citep{ingersoll1969,mueller1970}. The water vapour present in the current atmosphere may be re-supplied by chemical interaction with water-containing rocks \citep{mueller1970} or could be provided by meteoritic and cometary infall \citep{lewis1974}.

The present-day Earth hosts abundant water in all three states of matter --- solid (ice), liquid, and gas --- due to a fortuitous combination of surface temperature and pressure. The origin and evolution of water on Earth (and on the other terrestrial planets) is a subject of ongoing research. Two key questions are (1) whether Earth incorporated a sufficient amount of water at the time of accretion to explain its present day water content (`wet accretion') or if the accreted material was depleted in volatiles due to the high temperature in the inner solar nebula (`dry accretion'), and (2) what fraction of the Earth's water was delivered later by exogenous sources, e.g., comet and asteroid impacts \citep{drake-righter2002}. Comparing isotope ratios of volatiles in Earth, comets, asteroids, and meteorites, especially of the D/H ratio, can give us clues to answering these questions. However, it is not clear to what extent the D/H ratio in Vienna Standard Mean Ocean Water (VSMOW; the most commonly used standard isotopic reference for `Earth water' -- \citealt{Balsiger1995}) represents that of the early Earth. An increase of D/H by a factor 2--9 over the lifetime of the Earth due to mass fractionation during atmospheric loss is possible \citep{genda-ikoma2008}, and the Earth's lower mantle (supposedly least affected by atmospheric processes) has a D/H ratio lower by up to 20\% than VSMOW \citep{hallis-huss2015}. This places the Earth's D/H ratio between that of the protosolar nebula and that of comets from the outer Solar System (see, e.g., \citealt{saal-hauri2013,altwegg15} and references therein, and Section \ref{sec:d2h} below).

On Mars, water currently is present in the form of ice at the polar caps \citep{bibring-langevin2004,langevin-poulet2005} and in craters \citep{armstrong-titus2005,brown-byrne2008}, in a small amount of vapour in the atmosphere \citep{clancy-grossman1992}, and embedded in hydrated minerals \citep{bibring-langevin2006}. Liquid water may exist under specific conditions \citep{malin-edgett2006,martin-zorzano2015}. There is strong geologic and mineralogic evidence that liquid water was more abundant on Mars in the past, when the atmosphere was thicker \citep{sagan-toon1973,lasue-mangold2013}. Like on Venus, an elevated D/H ratio indicates that a significant amount of water vapour escaped to space from the atmosphere, affecting H$_2$O more than D$_2$O \citep{encrenaz-dewitt2016}. The original water content of Mars may have been sufficient to cover the planet with a layer of up to 1 km depth \citep{lasue-mangold2013}. There is currently no evidence for the existence of water on the Martian moons, Phobos and Deimos \citep{rivkin-brown2002}.

Saturn and Jupiter contain water in liquid and solid form in their lower cloud layers \citep{niemann-atreya1998,baines-delitsky2009}, but its abundance in these gas giants is not well known \citep{atreya-wong2005}. Uranus and Neptune are thought to contain a large layer of ices, including H$_2$O, above a rocky core. Also their atmospheres contain H$_2$O \citep{podolak-weizman1995}. The rings of the giant planets contain water ice at various fractions. While Saturn's rings consist mainly of water ice with a small admixture of organics and other contaminants \citep{nicholson-hedman2008}, the rings of Jupiter, Uranus and Neptune contain at best a small fraction of water ice \citep{lane-west1989,wong-depater2006,dekleer-depater2013}.  Many of the moons of the outer planets and Pluto contain a significant fraction of water, with Tethys possibly consisting almost entirely of water ice \citep{thomas-burns2007}. Some, such as Jupiter's Europa and Saturn's Enceladus, are thought to contain a tidally heated global or local subsurface ocean of liquid water beneath the outer shell of ice \citep{carr-belton1998,manga-wang2007,nimmo-spencer2007,hansen-esposito2008}.

\subsection{Kuiper Belt Objects, Centaurs and Comets}
Compared to the major planets in the Solar System, small icy bodies have experienced much less thermal evolution and their physical properties are well preserved.  The Kuiper Belt objects (KBOs) are numerous small icy bodies beyond Neptune, which are thought to be the most primitive remnants from the early Solar System. As these objects were formed far beyond the `snow line', volatile ices (especially water ice) are believed to be a principle constituent in KBOs. Optical and near-infrared spectroscopy is a widely used method to investigate surface properties as well as the chemical compositions of small bodies, although it is technically challenging to obtain spectral data on KBOs because of their great distance from the Sun and thus faintness.  For atmosphere-less bodies, evidence for water comes from solid-state absorption features in the spectrum of sunlight reflected from their surfaces (see section \ref{sec:water-obs}).  So far, spectra of KBOs fall into three categories, namely water-rich, methane-dominated and featureless ones \citep{Trujillo2011}.  For example, water ice was detected on KBOs (50000) Quaoar and Haumea (including its satellites) with two strong absorption bands at 1.5 $\mu$m and 2.0 $\mu$m respectively \citep{jewitt:2004, trujillo:2007, schaller:2008}. Other large KBOs  exhibit these bands as well as the absorption feature of crystalline ice at 1.65 $\mu$m, for example: Charon \citep{Brown:2000, Grundy:2016}, and Orcus \citep{Barucci:2008}. Haumea's satellite Hi'iaka and collisional family members are also known to be covered by crystalline ice \citep{Dumas:2011,schaller:2008,snodgrass-haumea,carry-haumea}. The spectra of Pluto, Eris and Makemake show a series of distinct bands of methane in the NIR \citep{Grundy:2016, brown:2005,licandro:2006}. Some trace elements, i.e. ammonia and methanol, have been detected on mid-sized KBOs \citep{Brown:2000, Barucci:2008, Barucci:2011}. Compared to spectroscopic observations, \cite{Trujillo2011} showed that a new near-infrared photometric system is sensitive to water ice and methane ice while reducing telescope observing time by a factor of $\sim$3. This system is particularly useful for surveying a large number of objects with moderate amount of telescope time. 

\begin{figure}
\includegraphics[width=\columnwidth]{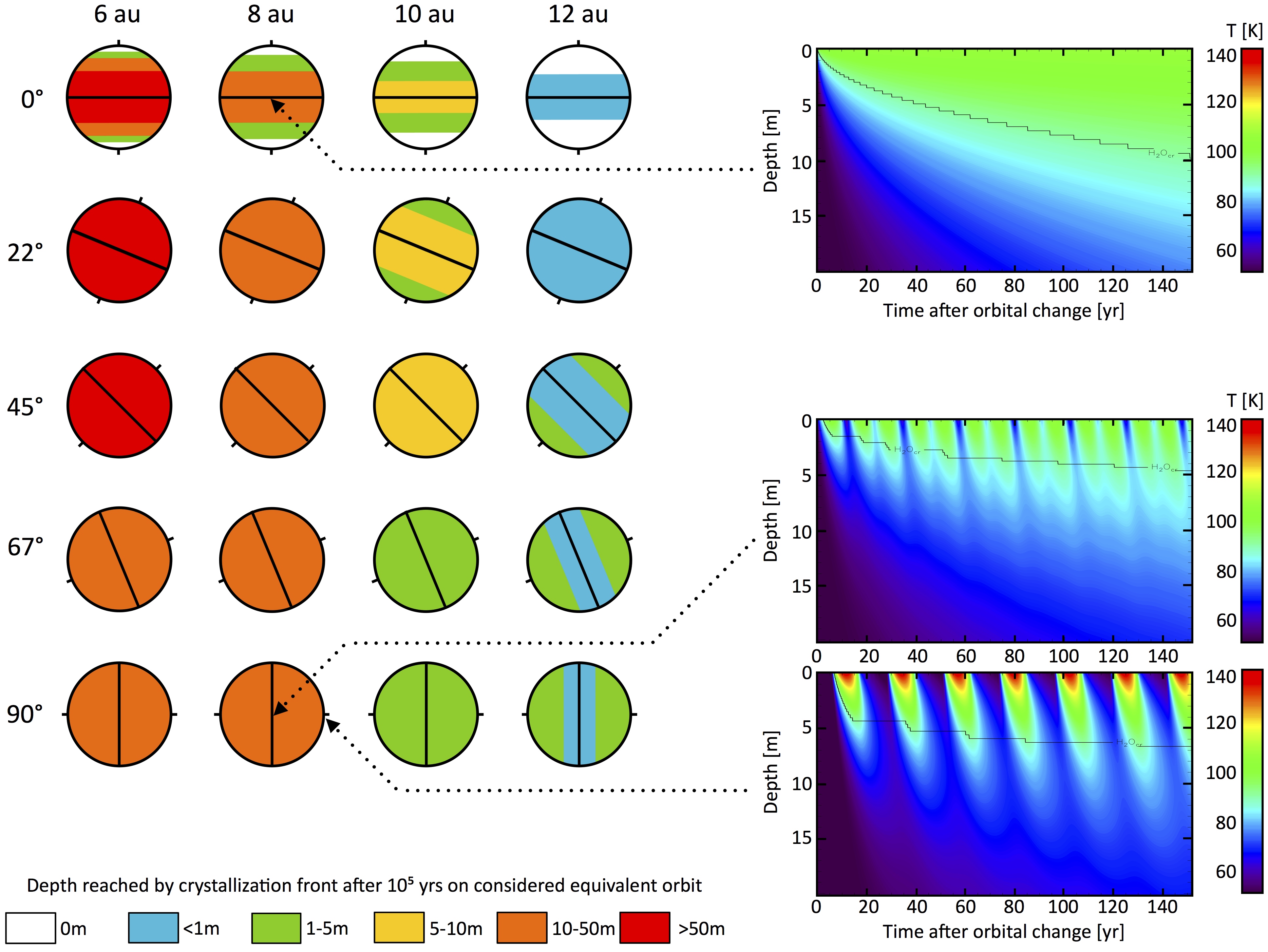}
\caption{{\it Left:} The depths where the crystallisation front stops for different [$a_c$ = equivalent semi major axis, obliquity] configurations, after $10^5$ years. Note that crystallisation may not be complete above, in the subsurface layer. The rotation period is very small compared to all other time scales involved.
{\it Right:} How the heat wave and crystallisation front progress in the subsurface layer for three cases (at 8 au and from top to bottom, under the equator at 0$^\circ$ obliquity, under the equator at 90$^\circ$ obliquity and under the pole at 90$^\circ$ obliquity), during the time the objects may be active ($\sim$ few $10^2$ years after an orbital change labelled time = 0). Black line shows the crystallisation fronts. From \citet{Guilbert-Lepoutre2012}.}
\label{fig:centaurs}
\end{figure}

According to \cite{jewitt:1998}, Centaurs are defined as objects with perihelia $q>a_J$ ($a_{J}=5.2$~au) and semimajor axes $a<a_N$ ($a_{N}=30$ au). These objects are widely believed to be `refugees' from the Kuiper belt \citep{levison:1997}, located on unstable orbits between Jupiter and Neptune with short dynamical lifetimes of about 10$^6$ to 10$^7$ years \citep{Dones:1999, horner:2004}.  Both optically blue and red members have been found among Centaurs \citep{tegler:2008, peixinho:2012, fraser:2012}.  (5145) Pholus is found to be one of the reddest objects observed to date in the Solar System \citep{fink:1992, davies:1993}, whose spectrum shows not only strong water ice bands but also an absorption complex at 2.27 $\mu$m \citep{cruikshank:1998}. \cite{dalle-ore:2015} studied seven TNOs and three centaurs that are among the reddest known. They conclude that these `ultra-red' objects in general might contain methanol/hydrocarbon ices and their organic materials could have been produced by irradiation of the volatile ices. \cite{jewitt09} reported observations of a sample of 23 Centaurs and found nine to be active. They found that `active Centaurs' in their sample have perihelia systematically smaller than the inactive ones. Centaurs have their perihelia beyond the water ice sublimation critical distance of $\sim$5 au, which suggests that the comet-like activity of Centaurs is driven by a mechanism different from water ice sublimation. Therefore, the cometary activity might be powered by conversion of amorphous ice into the crystalline phase and the subsequent release of trapped gases, such as carbon monoxide and carbon dioxide \citep{jewitt09}. Thermal evolution models have been used to study the occurrence of crystallisation, the depth that the front would reach, and how it can contribute to Centaurs activity for various orbits and obliquities (fig. \ref{fig:centaurs}; \citealt{Guilbert-Lepoutre2012}).

Comets generally have small, irregularly shaped nuclei that are composed of refractory materials (such as carbon, silicates etc) and icy grains (e.g., water ice, carbon dioxide ice and methanol ice).  These objects may have experienced very little alteration since their formation in the coldest regions of the young Solar System. As such, comets serve as the best probe for studying the physical conditions (e.g.,  temperature, pressure and composition) of the outer Solar nebula.  
However, direct detection of water ice in comets is surprisingly scarce. To date, solid ice features have only been observed in a handful of comets via ground-based facilities \citep{Davies:1997, Kawakita:2004,Yang:2009,Yang:2014} and {\it in situ} observations \citep{Sunshine:2006, AHearn:2012, Protopapa:2014,DeSanctis2015}.
Knowledge of ices in comets mainly comes from spectroscopic observations of emission lines and bands from the gas coma, which are discussed in more detail in section \ref{sec:water-obs}.

\subsection{Evidence for water in Asteroids}
 Although the asteroid belt could be an important source for Earth's water \citep{morbidelli:2000}, thermal models have shown that surfaces of most asteroids are too hot for water ice to remain stable against sublimation (see \citealt{schorghofer08} and section \ref{sec:ice-survival}). To date, only a few detections of surface ice in asteroids have been reported. On the other hand, hydrated minerals have been widely observed in asteroids \citep{rivkin:2015}. 
 
The largest body in the asteroid belt, (1) Ceres, contains significant water ice. Although its surface does not show signatures of water ice, the shape of Ceres was explained with a significant ice mantle \citep{Thomas2005}. Observations in the UV hinting at escaping water \citep{AHearn+Feldman1992} were confirmed in the sub-mm with detections with the Herschel space telescope \citep{Kueppers-ceres}. 
Using nuclear spectroscopy data acquired by NASA's Dawn mission, \cite{prettyman:2017} analyzed the elemental abundances of hydrogen, iron, and potassium on Ceres. They found that the regolith, at mid-to-high latitudes, contains high concentrations of hydrogen, consistent with broad expanses of water ice. The presence of ice in the largest asteroid has inspired debate about its origin, including suggestions that Ceres could have formed in the KBO region \citep{McKinnon2012}, but there is growing evidence that it is not the only icy asteroid, and that smaller bodies also have ice.

In the $\sim$100 km diameter range, asteroid (24) Themis  bears some significance in the debate of whether ice can originate from and survive in the Main Belt (MB), as water ice may have been detected at its surface \citep{campins10,rivkin10} via an absorption feature near 3 $\mu$m. We note that Themis was not found to be active by \citet{jewitt12a}, who suggested that if indeed present at its surface, water ice should be relatively clean and confined to a limited spatial extent, which was then confirmed by \citet{mckay17}.  \citet{jewitt12a} observed Themis far from perihelion, however, which as discussed in Section~\ref{sec:energybalance}, may mean that the sublimation strength at the time was several orders of magnitude weaker than might be expected for Themis closer to perihelion.  Many members of the Themis family show evidence for hydration \citep{florczak99,takir12}, and the second largest member of the family after Themis itself -- (90) Antiope -- might have some surface water ice \citep{hargrove15}. 
 
In a wider survey of large asteroids, \cite{takir12} identified four 3 $\mu$m spectral and orbital groups, each of which is presumably linked to distinct surface mineralogy. Searches for water and OH features in near-Earth objects (NEOs) were upended by the
discovery of OH in the lunar regolith by three spacecraft \citep{Sunshine:2009,pieters:2009,clark:2009}. Detections of water and/or hydroxyl on large asteroids have been reported for (175706) 1996 FG3 \citep{rivkin:2015}, (16) Psyche \citep{Takir:2017}, and (433) Eros and (1036) Ganymede \citep{Rivkin2017}. Although smaller (km-scale) asteroids are generally too faint for spectroscopy in the 3 $\mu$m region, the activity of MBCs offers evidence that water is also present in at least some smaller asteroids.


\section{The Main Belt Comets}\label{sec:MBCs}

\subsection{Definitions: Active Asteroids and Main Belt Comets}

An `active' asteroid can be thought of as any body in an asteroidal (rather than cometary) orbit which is observed to lose mass, normally through the observation of a dust tail or trail.  `Asteroidal' orbits have Tisserand parameters \citep{kresak82,kosai92} with respect to Jupiter, 
\begin{equation}
T_J = \frac{a_J}{a} + 2 \ \sqrt[]{\frac{a(1-e^2)}{a_J}} \cos(i) \geq 3.05, 
\end{equation}
and semi-major axes $a$ less than that of Jupiter ($a_J = 5.2$ au). Here $a$, $e$, and $i$ are the semimajor axes (in au), eccentricity, and inclination (in degrees) of the orbit, respectively. Such orbits are distinct from comets (which have $T_J<3.0$; cf. \citealt{levison96}). We use $T_J = 3.05$ instead of $T_J = 3.00$ because small drifts in $T_J$ beyond 3.0 are possible due to non-gravitational forces or terrestrial planet interactions \citep[e.g.,][]{levison2006,hsieh16_tisserand}. Other authors \citep[e.g.,][]{jewitt15} have used $T_J = 3.08$ as the boundary for similar reasons; the resulting list of MBC candidates is the same.

For the purposes of this paper, we use the term MBC to refer to an active asteroid that exhibits activity determined to likely be due to sublimation (e.g., from dust modeling results, or confirmation of recurrent activity near perihelion with intervening periods of inactivity). Various processes can be the source of mass loss for active asteroids in general, but repeated mass loss over several perihelion passages is most plausibly explained by sublimation. In addition, mass loss sustained over a prolonged period, although not unique to sublimation, can also be suggestive of ice sublimation since it is difficult to explain by a single impact, for example.

We are specifically not considering objects in the main asteroid belt whose cometary appearance has been shown to be due to mechanisms other than sublimation. These include debris released by impacts (either cratering events or catastrophic disruption) --- e.g., (596) Scheila \citep{bodewits11,jewitt11_scheila,Ishiguro11a,Ishiguro11b,moreno11_scheila} and P/2012 F5 (Gibbs) \citep{stevenson12,moreno12_2012F5} --- and rotational disruption --- e.g., 311P \citep{jewitt15_311P} --- thought to be an outcome of YORP spin-up for small asteroids \citep{scheeres15}. There are other hypothesized effects that have yet to be conclusively demonstrated to explain observed `activity', such as thermal cracking, electrostatic levitation of dust, or radiation pressure accelerating dust away from the surface \citep{jewitt15}.  We also exclude dynamically asteroidal objects (i.e., with $T_J>3.05$) outside of the asteroid belt for which activity has been detected, e.g., (3200) Phaethon \citep{jewitt10,li13,hui17_phaethon} and 107P/(4015) Wilson-Harrington \citep{fernandez97}. See \citet{jewitt15} and references therein for more extensive discussion of active asteroids beyond our definition of `MBC'. 
MBCs are of special interest within the broader class of active asteroids as they indicate the possible presence of water in bodies of a size that are very common in the asteroid belt, implying that there is potentially a large population of icy bodies there. 

\begin{table}
\caption{MBC discovery circumstances}
\label{tab:mbcdiscovery}       
\begin{tabular}{llllr}
\hline\noalign{\smallskip}
Object & Discovery date & Tel.$^a$ &  $m_V$$^b$ & \multicolumn{1}{c}{$\nu$$^c$} \\
\noalign{\smallskip}\hline\noalign{\smallskip}
133P/Elst-Pizarro              & 1996 Jul 14 & ESO 1.0m  & 18.3 & 21.6$^{\circ}$  \\
238P/Read                      & 2005 Oct 24 & SW 0.9m   & 20.2 & 26.4$^{\circ}$  \\
176P/LINEAR                    & 2005 Nov 26 & GN 8.1m   & 19.5 & 10.1$^{\circ}$  \\
259P/Garradd                   & 2008 Sep 2  & SS 0.5m   & 18.5 & 18.5$^{\circ}$  \\
324P/La Sagra                  & 2010 Sep 14 & LS 0.45m  & 18.3 & 20.1$^{\circ}$  \\
288P/2006 VW$_{139}$           & 2011 Nov 5  & PS1 1.8m  & 18.7 & 30.7$^{\circ}$  \\
P/2012 T1 (PANSTARRS)          & 2012 Oct 6  & PS1 1.8m  & 19.6 & 7.4$^{\circ}$   \\
P/2013 R3 (Catalina-PANSTARRS) & 2013 Sep 15 & PS1 1.8m  & 20.5 & 14.0$^{\circ}$  \\
313P/Gibbs                     & 2014 Sep 24 & CSS 0.68m & 19.3 & 8.0$^{\circ}$   \\
P/2015 X6 (PANSTARRS)          & 2015 Dec 7  & PS1 1.8m  & 20.7 & 328.9$^{\circ}$ \\
P/2016 J1-A/B (PANSTARRS)      & 2016 May 5  & PS1 1.8m  & 21.4 & 345.9$^{\circ}$ \\
\noalign{\smallskip}\hline
\end{tabular}
\smallskip
\newline $^a$ Discovery telescope: CSS 0.68m: Catalina Sky Survey 0.68m; ESO 1.0m: European Southern Observatory 1.0m; GN 8.1m: Gemini-North 8.1m; LS 0.45m: La Sagra 0.45m; PS1: Pan-STARRS1 1.8m; SS 0.5m: Siding Spring 0.5m; SW 0.9m: Spacewatch 0.9m
\newline $^b$ Approximate mean reported $V$-band magnitude at time of discovery
\newline $^c$ True anomaly in degrees at time of discovery
\end{table}

\subsection{Discovery of MBCs}

Since the discovery of 133P/Elst-Pizarro in 1996, the first recognised MBC, as of September 2017 ten more members have been detected. We summarise discovery circumstances of the currently known MBCs in Table~\ref{tab:mbcdiscovery}. 
The discovery rate has increased significantly since 133P was discovered, thanks to wide-field sky surveys, especially those designed to find transients or NEOs. Two new MBCs have been added to the list since the review by \citet{jewitt15}: P/2015 X6 (PANSTARRS), and P/2016 J1 (PANSTARRS), both discovered by the Pan-STARRS survey at Haleakala, Hawaii. P/2016 J1 is particularly interesting as it was observed to split into two pieces \citep{hui17_16J1,moreno17_16J1}, and its current activity perhaps was triggered by a small impact, whereby formerly buried ice started sublimating and the torque rapidly drove the parent to fragment \citep{hui17_16J1}. It reinforces the idea that rotational instability is one of the important fates that comets may suffer, especially for those sub-kilometre sized. In addition to sky surveys such as Pan-STARRS, there have been attempts to search for MBCs with targeted observations of known asteroids in main-belt orbits to search for activity \citep{hsieh09_htp}, and in archival data from the Canada-France-Hawaii Telescope \citep{Gilbert2009,Gilbert2010,Sonnett2011} and the Palomar Transient Factory \citep{Waszczak2013}. Recently, a citizen science project\footnote{\url{http://www.comethunters.org}} to crowd source the visual identification of MBCs has been employed in the search \citep{hsieh16_comethunters,comethunters}.  To date none of these targeted searches have discovered any MBCs.

\subsection{Population estimates}

Based on Pan-STARRS1 discovery statistics, on the order of $\sim$50--150 currently active MBCs comparable in brightness to the known MBCs are expected to exist \citep{hsieh15_ps1}, where their respective individual activity strengths will depend on their orbit positions at the time of observation.  We have not yet found them all as they need to be both active (which they are for only a few months of a 5--6 year orbit) and well placed for observations by surveys at the same time. While the pace of MBC discoveries has accelerated somewhat relative to the past to roughly one per year since the start of the Pan-STARRS1 survey (Table~\ref{tab:mbcdiscovery}), an even higher rate of discoveries will be necessary to increase the known population to the point at which it can be meaningfully statistically analyzed.

Assuming that we have mostly been discovering the brightest members of the MBC population, increasing the discovery rate will most likely require increasing the sensitivity of search efforts,  to find similar MBCs at less optimal observing geometries, or to sample the population of intrinsically fainter targets with lower activity levels, which are presumably more numerous.  To date, nearly all of the known MBCs have been discovered by telescopes with apertures smaller than 2 metres.  The use of larger telescopes for conducting surveys (e.g., the Dark Energy Survey's 4 m Blanco telescope or the 8.4 m Large Synoptic Survey Telescope) immediately allows for more sensitive searches for MBCs, but their respective observing efforts must also be coupled with at least reasonably capable search algorithms like that used by Pan-STARRS1 \citep[cf.][]{hsieh12_288p}.

Other than the direct confirmation of sublimation products from a MBC, the next most critical priority for advancing MBC research is the discovery of more objects.  A larger population of known objects (perhaps an order of magnitude larger than the currently known population) will allow us to achieve a more meaningful understanding of the abundance and distribution of MBCs in the asteroid belt as well as their typical physical and dynamical properties.

\begin{table}
\caption{Upper limits on MBC gas production (adapted from \citealt{snodgrass-T1})}
\begin{center}
\begin{tabular}{l l c S[table-format = 1.1e2] S[table-format = 1.1e2] l}
\hline
MBC & Tel. & $r$  & {$Q$(CN)}         & {$Q$(H$_2$O)}$^a$     & reference \\
    &      & (au) & {molec.\,s$^{-1}$} & {molec.\,s$^{-1}$} &           \\
 \hline
133P      & VLT      & 2.64 & 1.3e21 & 1.5e24 & \citet{licandro11_133p176p} \\
176P$^b$  & Herschel & 2.58 & {--}   & 4e25   & \citet{devalborro12}        \\
324P      & Keck     & 2.66 & 3e23   & 1e26   & \citet{hsieh12_324p}        \\
259P      & Keck     & 1.86 & 1.4e23 & 5e25   & \citet{jewitt09}            \\
288P      & Gemini   & 2.52 & 1.3e24 & 1e26   & \citet{hsieh12_288p}        \\
          & GTC      & 2.52 & 1.1e24 & {--}   & \citet{licandro13_288p}     \\
596$^c$   & Keck     & 3.10 & 9e23   & 1e27   & \citet{hsieh12_scheila}     \\
P/2013 R3 & Keck     & 2.23 & 1.2e23 & 4.3e25 & \citet{jewitt14_p2013r3}    \\
313P      & Keck     & 2.41 & 1.8e23 & 6e25   & \citet{jewitt15_313p}       \\
P/2012 T1 & Keck     & 2.42 & 1.5e23 & 5e25   & \citet{hsieh13_p2012t1}     \\
          & Herschel & 2.50 & {--}   & 7.6e25 & \citet{orourke13}           \\
          & VLT      & 2.47 & {--}   & 8e25   & \citet{snodgrass-T1}        \\
 \hline
\end{tabular}
\end{center}
$^a$ Where both $Q$(CN) and $Q$(H$_2$O) are given, the latter is derived from the former. Values quoted in the original reference are given.\\ 
$^b$ 176P was not visibly active (no dust release) at the time of the Herschel observations.\\
$^c$ The dust ejected from (596) Scheila was almost certainly due to a collision, rather than cometary activity \citep[e.g.,][]{bodewits11,Ishiguro11a,Ishiguro11b,Yang11}.
\label{tab:CNlimits}
\end{table}%

\subsection{Outgassing from MBCs}
\label{sec:outgassing}

The gas comae of comets are generally observed through fluorescence emission bands of various species, across a wide range of wavelengths from the UV to radio. 
The CN radical has strong emissions in the optical range, especially between 3590 and 4220 \AA. The (0-0) band at 3883 \AA\ is one of the most prominent feature of comet optical spectra. It has been detected at up to almost 10~au for comet Hale-Bopp \citep{rauer03} and it is usually one of the first gas emissions detected in comets from the ground. The CN radical can also be used as a proxy of water production. The ratio of water production rates to CN production rates in the coma of a typical Jupiter-family comet is around $Q$(H$_2$O)/$Q$(CN) = 350, even though variations of the ratio have been observed along the orbit of a comet as well as from one comet to another \citep{ahearn95}.

Upper limits of the CN production rate have been determined for several MBCs using spectroscopic observations on large telescopes: the Very Large Telescope (VLT), Gemini, the Gran Telescopio Canarias (GTC), and Keck. \cite{hsieh12_288p} and \cite{licandro13_288p} respectively determined upper limits of $Q$(CN) = $1.3\times 10^{24}$  and $Q$(CN) = $3.76\times10^{23}$~molec.\,s$^{-1}$ for 288P, \cite{hsieh13_p2012t1} report an upper limit of $Q$(CN) = $1.5\times10^{23}$~molec.\,s$^{-1}$ for P/2012 T1 (PANSTARRS), and \cite{licandro11_133p176p} measure an upper limit of $Q$(CN) = $1.3\times10^{21}$~molec.\,s$^{-1}$ for 133P. For comparison, we note that these upper limits are in line with the lowest recorded production rates for `normal' comets near 1~au; one of the lowest successfully measured was 209P/LINEAR with $Q$(CN) = $5.8\times10^{22}$~molec.\,s$^{-1}$ \citep{schleicher16}, while more commonly measured CN production rates are at least a few times 10$^{23}$ \citep{ahearn95}. The derived water production rates from these measurements, assuming a `typical' water/CN ratio for comets, are around $Q$(H$_2$O) = $10^{25} - 10^{26}$ molec.\,s$^{-1}$, which would be low for a typical Jupiter-family comet (JFC) at MB distances (e.g., Rosetta measurements for 67P inbound agree at $Q$(H$_2$O) $\sim 10^{26}$ molec.\,s$^{-1}$ at 3 au, and a few times this at around 2.5 au, where most of the MBC measurements were made; \citealt{Hansen2016}). Attempts to detect CN in MBCs to date are summarised by \citet{snodgrass-T1} and listed in table~\ref{tab:CNlimits}.

These upper limits on water production all rely on the assumption that MBCs have a similar water/CN ratio to other comets, which is unlikely to be the case (see detailed discussion in following sections). Attempts have also been made to directly detect outgassing water using the Herschel space telescope \citep{devalborro12,orourke13}, and to detect the photo-dissociation products of water (OH and O) with the VLT \citep{snodgrass-T1,Jehin2015}. These are discussed in more detail in section \ref{sec:water-obs}. A second implicit assumption in any upper limit measurement is that gas is still present at the time of the observation (i.e. that activity is ongoing). In most cases these observations were performed when there was dust visible around the MBC, and ongoing activity is a reasonable assumption, but it is also possible that an initial period of activity lifted dust and then shut off, leaving slow moving dust coma and tail visible after faster moving gas already dispersed. Models of dust coma morphology can be used to differentiate between ongoing activity and remnant dust from short impulsive events \citep[e.g.,][]{moreno11_324p}.


\section{MBC origins and survival of ice}\label{sec:ice-survival}

Constraining the origin of MBCs is particularly important in the context of understanding whether these objects can be representative of ice-rich asteroids native to the MB, or have been implanted from outer regions of the Solar System during its  complex dynamical evolution \citep[cf.][]{hsieh14_prociaus}.  MBCs could be representative of the source of the terrestrial planets' volatiles and of the Earth's oceans in particular, and can therefore be extremely astrobiologically significant. 
It is also important to understand to what degree MBCs can be expected to have compositions which are comparable to traditional comets, e.g., their H$_2$O/CN ratio.

\subsection{Isotopic ratios}
\label{sec:d2h}

Isotopic ratios, and particularly the D/H ratio in water, are used to trace the formation location of Solar System ices. In the protoplanetary disc, reactions in the vapour phase meant that D/H varied with temperature, but when the ice froze out, the ratio became fixed, meaning that the D/H ratio observed now records, in some way, the temperature and therefore location at which the ice originally formed. D/H is expected to increase with the heliocentric distance at which a body formed \citep{Robert2000,Robert2006}. As mentioned in Section \ref{sec:ice-everywhere}, the Earth's oceans have a mean D/H value that is lower than that measured in most comets, including the most recent measurement by Rosetta for 67P \citep{mumma11,altwegg15}. As the D/H values measured in meteorites are closer to VSMOW, it is possible that asteroids are instead the dominant source of Earth's water. Therefore, ice-rich asteroids such as MBCs are potentially remnants of the population that supplied our oceans. However, no D/H ratio has yet been measured for a MBC, and a space mission to visit one would be required to do this (see Section \ref{sec:future}). This is seen as a priority for future measurements, both to confirm the match between ice in MBCs and Earth's water, but also to better understand the original source region where MBCs formed.
 In addition to D/H, isotopic ratios in other volatile elements (O, C, N, S) are measured in comets \citep{Jehin2009,Bockelee-Morvan2015}. Combining information from different elements can place stronger constraints on source location within the protoplanetary disc, but these measurements are even more challenging than D/H, given the relatively low abundance of other volatile species relative to water.

\subsection{Dynamical evolution}

Studying the dynamical stability and evolution of MBCs is a key component in efforts to identify their likely source regions in the Solar System. Most MBCs have been found to be mostly stable over 10$^8$ years or more \citep{jewitt09,haghighipour09,hsieh12_288p,hsieh12_324p,hsieh13_p2012t1}, suggesting that they formed {\it in situ} where we see them today in the asteroid belt.  These results were corroborated by \citet{hsieh16_tisserand} who studied synthetic test particles rather than real objects, concluding that objects on orbits with both low eccentricities and low inclinations are unlikely to have been recently implanted outer Solar System objects.

Despite the fact that most MBCs appear to occupy long-term stable orbits, some MBCs like 238P/Read and 259P/Garradd are rather unstable on timescales of the order of 10$^7$ years \citep{jewitt09,haghighipour09}, suggesting that they may be recently emplaced.  Though the architecture of the modern Solar System appears to largely lack reliable pathways by which outer Solar System objects can evolve onto MBC-like orbits \citep[e.g.,][]{fernandez02}, such interlopers cannot be completely excluded.  Considering only the dynamical influence of major planets and the Sun, \citet{hsieh16_tisserand} showed that pathways exist that are capable of temporarily implanting  JFCs in the MB, largely shaped by the effects of close encounters with the terrestrial planets.  However, they also showed that such implanted objects would not be stable for more than 100~Myr.
These dynamical pathways appear to work both ways, with a non-negligible possibility that some JFCs could originate from the MB \citep{jfernandez15}.

A final scenario to consider is whether planet migrations during the early stages of Solar System evolution, as described by the Grand Tack and Nice models, could have resulted in the emplacement of icy outer Solar System objects in the MB \citep{levison09,walsh11}. In this case, ice formed in the outer regions of the Solar System could have been implanted in the MB, and would have experienced a different thermal history from ice that remained in the outer Solar System, thus allowing us to probe processes related to the thermophysical evolution of icy bodies in general.  As discussed earlier in this paper though, much work remains in developing and performing observational tests for these models as well as refining them to the point where predictions (e.g., the distribution in orbital element space of implanted objects in the MB) can be made and tested. 

\cite{hui16} is the first work which systematically examines nongravitational effects of the MBCs. The authors report statistically significant detections of nongravitational effects for 313P/Gibbs and 324P/La Sagra. Intriguingly, the nongravitational effect on 324P is found to be large (composite nongravitational parameter $\sim$10$^{-7}$ au day$^{-2}$; \citealt{hui16}), which may be correlated with the fact that it is one of the most active MBCs \citep{hsieh14_324p}, and may support the argument that some MBCs may have originated as JFCs \citep{hsieh16_tisserand}. For the rest of the MBCs, \cite{hui16} fail to obtain meaningful detections on nongravitational effects, which is consistent with the fact that the activity of the MBCs is normally orders-of-magnitude weaker than typical comets. Given this, conclusions about dynamical evolutionary paths of the MBCs by previous authors without consideration of the nongravitational effects are likely unaffected.

\subsection{Family origins}

In addition to the dynamical considerations discussed above, we need to account for the formation of asteroid families by catastrophic disruption of large parent bodies when trying to constrain the origins of MBCs. MBCs P/2012 T1 and 313P have for example been linked to the Lixiaohua family \citep{hsieh13_p2012t1,hsieh15_313p}, while 133P, 176P/LINEAR, 238P and 288P/2006 VW$_{139}$ have been linked to the Themis family \citep{toth00,hsieh09_htp,hsieh09_238p,hsieh12_288p}.  MBC 324P has also been dynamically linked with a small cluster of six objects (including 324P itself), but due to the small number of cluster members identified to date, this grouping does not yet formally satisfy the criteria for being classified as a family or clump as defined by \citet{novakovic2011_highifamilies}.

The link between several MBCs and the Themis family is interesting, not only because of the water ice detection on Themis itself, but because
spectra that have been obtained for Themis asteroids are best matched by carbonaceous chondrites meteorites with different degrees of aqueous alteration \citep{fornasier16,marsset16}. These observations are consistent with a large parent body made of significant amounts of water ice, which differentiated (hence the hydration of minerals) but maintained layers of pristine/unheated material \citep{castillo-rogez10}. The layering of composition, or internal heterogeneity, could in this case explain the spectral variability observed among Themis family members, which cannot be completely explained by space weathering processes \citep{fornasier16}. Among the Themis family lie much younger sub-families: the Beagle family, aged less than 10~Myr \citep{nesvorny08}, and the 288P cluster, estimated to be 7.5~Myr old \citep{novakovic2012_288p}. It is possible that MBCs 133P and 288P may be members of those young sub-families rather than primordial members of the Themis family, increasing the possibility that water ice -- if initially present in the various precursors -- could have survived in both objects until today.

We note that while young families similar to those found for 133P and 288P have not yet been formally identified for the other known MBCs, the small sizes of other MBC nuclei suggests that they too may be fragments of recent catastrophic disruption events.  Smaller objects are collisionally destroyed on statistically shorter timescales relative to larger objects \citep{cheng2004_collisions,bottke2005_collisions}, meaning that currently existing smaller objects are statistically more likely to have been recently formed (e.g., in the fragmentation of a larger parent body) than larger bodies.  The young families resulting from those fragmentation events may perhaps simply have not yet been identified because an insufficient number of members have been discovered to date, preventing the identification of the families by standard clustering analyses.  As more asteroids are discovered by ongoing and future surveys, it will therefore be useful to periodically re-run clustering analyses for all MBCs to check if any new young families can be identified.

\subsection{Thermal processing and ice survival}

\citep{hsieh15_ps1} noted that MBC activity patterns are predominantly modulated by variations in heliocentric distance. This observation implies that such activity is the result of ice being present on a global scale, buried under a slowly growing dust mantle, allowing for activity to be sustained over multiple perihelion passages, rather than the result of isolated local active sites containing exposed ice being seasonally illuminated (which would tend to produce activity more randomly distributed along object orbits). However, given the small number of MBCs and the different individual cases, both scenarios have to be taken into account. \citet{schorghofer08} studied the survival of water ice inside asteroids, assumed to be spherical icy objects on orbits ranging from 2 to 3.3~au from the Sun. They introduced the concept of the `buried snowline', i.e. the limit beyond which subsurface water ice can be sustained for the age of the Solar System inside asteroids. Due to the very low thermal conductivities observed for such objects, they found that ice could have survived for billions of years in the top few metres from the surface, provided the mean temperature of the surface remains below 145~K: this is achievable in the polar regions of most objects in the outer MB with low obliquity.

Individual studies have mainly focused on 133P, since it has been active for four consecutive perihelion passages.  \citet{prialnik09} studied two aspects of the survival of ices in 133P, both during the long-term thermophysical evolution of the object since its formation, and after an impact in the recent history of the object. For the 4.6~Gyr evolution, they used a one-dimensional thermophysical model able to compute both the thermal history of the object and the retreat of various volatiles, in particular, water, CO, and CO$_2$. They assumed that 133P was formed in the outer Solar System and then implanted in the MB, where the larger equilibrium temperature would have caused ices to sublimate, causing the sublimation front to penetrate deeper below the surface. They found that only water ice would have been able to survive inside 133P, 50--150~m below its surface. Other minor species would have all been lost, and would not have been sustained even in the case of an extremely low thermal conductivity. From these results, they infer that in order for 133P to be active today, an impact must have occurred, by which material was removed from the surface to expose ice-rich layers. This idea is consistent with the study of \citet{capria12} who found that impact rates in the MB are consistent with MBC activity being the result of impacts able to expose water ice at the surface of asteroids. 

With this scenario in mind, \citet{prialnik09} then studied the thermal processing of 133P with a fully three-dimensional model which allows the assessment of the latitudinal variations of sublimation and dust mantling. Their results show that a dust mantle is being slowly built after repeated perihelion passages, with a thickness that is not uniform across the surface due to latitudinal variations of ice evolution. It is interesting to note that their analysis includes results consistent with the main features observed in 133P's secular lightcurve as studied by \citet{ferrin06}, in particular an increase in brightness of $\sim$2~mag above the bare nucleus at peak activity, and a time lag of $\sim$150 days after perihelion passage for peak activity. \citet{prialnik09} were able to reproduce these features when considering large tilt angles in their simulations. A large obliquity was suggested for 133P \citep{toth06}, but has not been confirmed \citep{hsieh10_133p}. This would indicate that the behavior expected from water ice evolving in the MB can reproduce the activity pattern of this MBC: this implies that although no signatures of gaseous species have been detected, ice can be the origin of MBC activity, at least for 133P.

\begin{figure}
\includegraphics[width=\columnwidth]{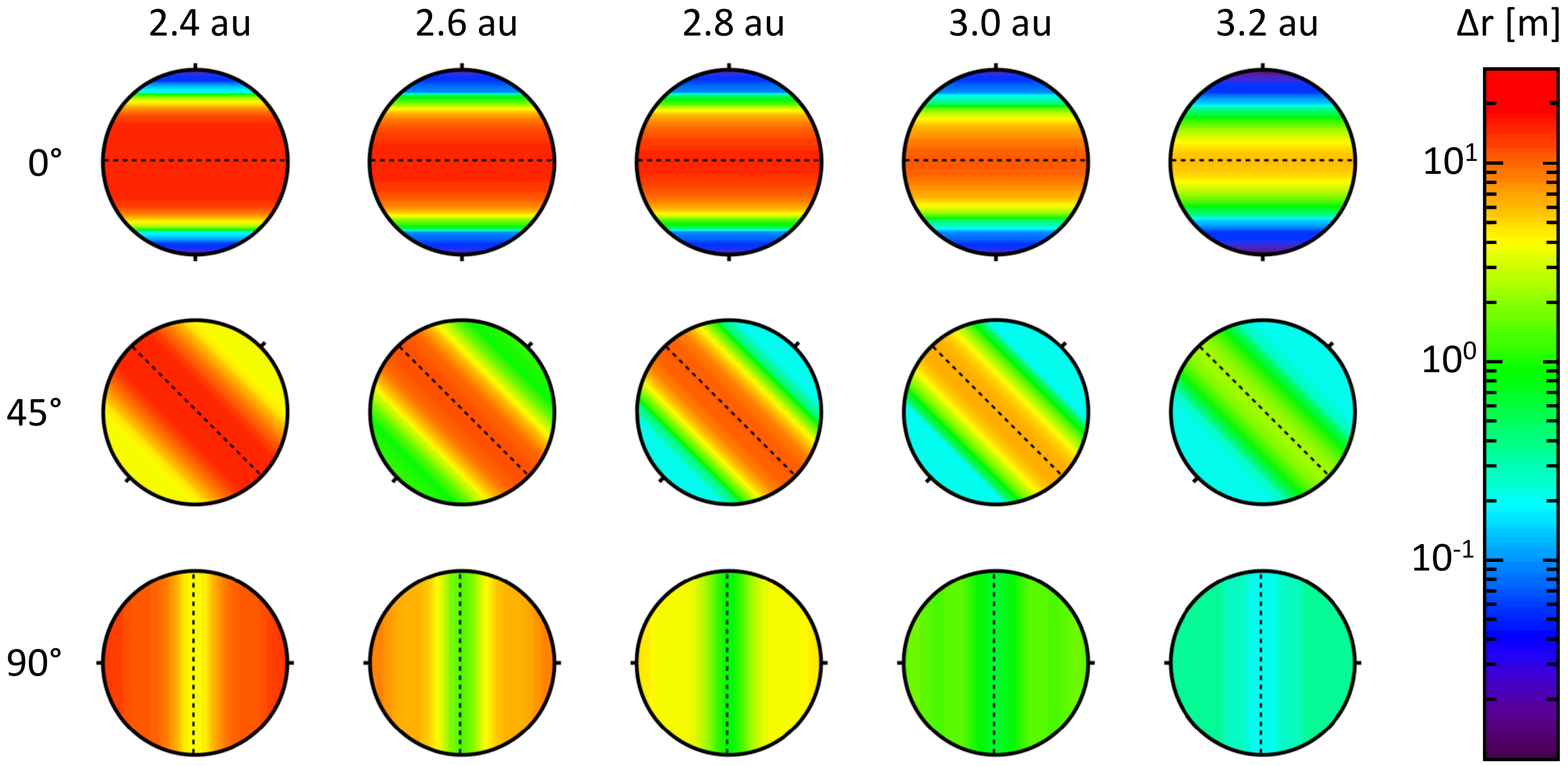}
\caption{Thickness of the porous dusty crust at the surface of MBCs required for water ice to be thermodynamically stable underneath after 1~Myr on equivalent orbits $a_c = a(1 - e^2)$ displayed as columns. Each row corresponds to a tilt angle between the object's rotation axis and its orbital plan. The Bond albedo is 2\% and the thermal inertia at the surface is 3~J~K$^{-1}$~m$^{-2}$~s$^{-1/2}$.}
\label{fig:thermal-model}
\end{figure}

\begin{table}
\caption{MBC properties relevant for thermal models$^a$ 
}
\label{tab:mbcparamaters}       
\begin{tabular}{lclcccc}
\hline\noalign{\smallskip}
Object & $a$$^a$ & $e$$^b$ & $a_c$$^c$ & E$^d$ & $T_q$$^e$  \\
\noalign{\smallskip}\hline\noalign{\smallskip}
     133P & 3.166 & 0.159 & 3.08  & $4.65\times 10^{15}$ & 160    \\
     238P & 3.165 & 0.252 & 2.96  & $4.74\times 10^{15}$ & 163    \\
     176P & 3.196 & 0.193 & 3.07  & $4.66\times 10^{15}$ & 161    \\
     259P & 2.726 & 0.342 & 2.40  & $5.27\times 10^{15}$ & 175    \\
     324P & 3.096 & 0.154 & 3.02  & $4.69\times 10^{15}$ & 162    \\
     288P & 3.049 & 0.201 & 2.92  & $4.78\times 10^{15}$ & 164    \\
P/2012 T1 & 3.154 & 0.236 & 2.98  & $4.73\times 10^{15}$ & 163    \\
P/2013 R3 & 3.033 & 0.273 & 2.80  & $4.88\times 10^{15}$ & 167    \\
     313P & 3.156 & 0.242 & 2.97  & $4.74\times 10^{15}$ & 163    \\
P/2015 X6 & 2.755 & 0.170 & 2.67  & $4.99\times 10^{15}$ & 170    \\
P/2016 J1 & 3.172 & 0.228 & 3.00  & $4.71\times 10^{15}$ & 162    \\
\noalign{\smallskip}\hline
\end{tabular} 
\smallskip
\newline $^a$ Semimajor axis, in au 
\newline $^b$ Eccentricity
\newline $^c$ Semimajor axis, in au, of the equivalent circular orbit, i.e., the circular orbit receiving the same amount of energy per orbit
\newline $^d$ Insolation per orbit, in W m$^{-2}$
\newline $^e$ Theoretical equilibrium surface temperature at semimajor axis of  equivalent circular orbit, assuming a sublimating isothermal gray-body (see Section~\ref{sec:energybalance} for description of calculations)
\end{table}

Using a different modeling approach, namely an asynchronous model coupling the simulation of accurate diurnal variations of the temperature with the simulation of ice sublimation over long timescales, \citet{schorghofer16} revisited the thermo-physical evolution of 133P. In particular, \citet{prialnik09} focused on 133P having spent billions of years in the MB, when we know from dynamical studies that it belongs to the Themis family (aged 2.5 Gyr) and perhaps to the Beagle family (aged 10 Myr) and thus has only been a discrete object since breaking up from the parent body at those times. Assuming that 133P is a direct member of the Themis family, \citet{schorghofer16} found that water ice could indeed be found at depths consistent with those of \citet{prialnik09}, but only for large grain sizes in the polar regions of 133P, i.e. ice can be found 2.4 to 39~m below the surface for 1~cm grains. When considering smaller grain sizes, \citet{schorghofer16}'s results suggest that ice could be expected to be found much closer to the surface, and as close as 0.2~m for 100 $\mu$m grains. Assuming that 133P is a member of the much younger Beagle family, they suggest that ice could be found as close as a few cm below the surface (0.07 to 0.16~m for 100 $\mu$m grains) and within the top few metres in all cases (above 3.9~m for 1~cm grains). 

Such different results between the two studies can in part be due to the different modeling approaches, or the initial thermo-physical properties assumed in the models. A better understanding of the internal structure of MBCs is crucial, however, for assessing the expected production rates, whether minor volatiles could have survived, and therefore to prepare for ground or space-based observations which might confirm the presence of ice on MBCs. Although an in-depth study of the survival of water ice and other volatiles is beyond the scope of this paper, we choose to provide an additional input given the discrepancy between the main studies of 133P's internal structure and evolution.

To further explore the stability of water ice in the interior of MBCs, we used the method described in \citet{guilbert-lepoutre14} who studied the survival of water ice on Jupiter Trojans. Below the orbital skin depth ($\sim$a few to tens of metres), the material is not sensitive to variations of the surface temperature, but to the mean annual temperature. We therefore assumed equivalent circular orbits which receive the same amount of energy per orbit as asteroids, including MBCs, using $a_c = a(1 - e^2)$ as orbital input in a numerical model of three-dimensional heat transport to constrain the temperature distribution over the range of heliocentric distances spanned by MBCs (see Table \ref{tab:mbcparamaters}). We have also considered a tilt angle between the rotation axis and the MBC orbit plane of 0, 45 and 90$^\circ$ to obtain a limit on the dust layer thickness required for ice to survive underneath. For the main thermo-physical parameters, we have used a 2\% albedo and a low thermal inertia of 3~J~K$^{-1}$~m$^{-2}$~s$^{-1/2}$, which is consistent with values measured for comets, though among the lowest in the range of possible inertias (further details can be found in \citealt{guilbert-lepoutre14} and \citealt{guilbert-lepoutre11}). In Figure~\ref{fig:thermal-model} we show the depth at which water ice can be found after 1~Myr for the different orbital configurations considered, as preliminary results for a dedicated study of ice survival in the MB.  The water ice depth ranges from a few centimetres to about 30~m for objects as close to the Sun as $a_c=2.4$~au. It is worth noting that already after 1~Myr, highly volatile species such as CO or CH$_4$ would have already been lost. However, less volatile species such as CO$_2$, HCN or CH$_3$OH could still be present within the top 30--100~m, especially at large heliocentric distances. 
 Although these results cannot be directly compared with existing works in  meaningful way, given the different timescales simulated by the different approaches, they appear more consistent with the results obtained by  \citet{schorghofer08}  on the survival of water ice, and suggest that a detailed analysis needs to be performed for constraining the survival of minor volatile species.


\section{Detecting water}\label{sec:water-obs}

As discussed in section \ref{sec:outgassing} above, CN has been used to search for outgassing around MBCs, but without success.
To use MBCs as probes of the water content of the MB we need to directly detect water (or its daughter/grand-daughter species). In this section we discuss the various means by which this is achieved in comets, and the prospects for applying these techniques to MBCs.

There are several techniques and wavelength ranges where water production rates in comets are determined. Water molecules themselves can be detected in their vibrational fluorescence transition in the NIR at 2.7 $\mu$m, however, the main band emission cannot be detected from the ground because of strong absorption by telluric water in the atmosphere. There are some weaker hot bands in the IR that can be detected and are routinely observed in medium to bright comets and at heliocentric distances much smaller than those of MBCs \citep{bockele-morvan:2004}. Rotational lines of water for the same reason also cannot be observed from the ground but have been observed from space-borne platforms like ISO \citep{Crovisier1997} and SWAS \citep{Chiu2001}. This is also how the most sensitive detection of water was made on approach by the Rosetta spacecraft to comet 67P by the MIRO instrument, a small radio telescope \citep{Biver2015}.

Otherwise water production is normally determined with greater sensitivity by observing water dissociation fragments such as OH, H and O($^1$D).  The OH radical is the primary product of the photodissociation of water at the level of about 82\%. Its UV emission in the (0-0) band at 3080 \AA{} is observed both from ground-based and space-based telescopes \citep{Feldman04}. The fluorescence of the UV emissions produces a population inversion of rotational levels that are observed with radio telescopes at 18 cm \citep{despois1981,schleicher1988}.
Space-based observations of atomic hydrogen have also been used as a proxy for determining water production rates. The chain of H$_2$O and OH photodissociation produces two H atoms per water molecule with nascent velocities of $\sim$18 and $\sim$8 km s$^{-1}$, respectively \citep{keller1976}; measurements of the Lyman-$\alpha$ emission at 1216 \AA{} can be analyzed to determine water production.  
Oxygen can be used via the forbidden lines of O($^1$D) at 6300 \AA{} \citep{fink2009} in the inner coma, but the lowest production rates determined this way have been only a little lower than those from H Lyman-$\alpha$, on the order of a few $\times 10^{27}$ molecules s$^{-1}$ \citep{fink2009},  also for comets near a heliocentric distance of about 1 au and very small geocentric distances.

In table \ref{tab:comet-detections} we list the various emission features of water (and selected other key cometary species, for reference), along with example observations of these species in comets. We have tried to select the weakest detection, i.e. lowest $Q$(H$_2$O), to show the limits of current technology, but of course the possibilities with very different facilities vary over orders of magnitude. Further examples of observations in different wavelength regimes are given in more detail in the following sub-sections. We give the diameter of the telescope used and the signal-to-noise per second achieved in order to allow some comparison to be made -- we use this and the geometry of each observation to approximately scale the different capabilities to the expected MBC case in section \ref{sec:MBC-expected-levels}. It is worth noting that comet 67P is often the example chosen, as this otherwise faint comet was observed with all possible techniques as part of the campaign of observations supporting the Rosetta mission \citep{snodgrass2017}, while for many species observations were only possible for very bright or very nearby comets (e.g., Hale-Bopp or 103P/Hartley 2; the latter was also a popular target due to the fly-by of the NASA EPOXI mission close to perihelion -- \citealt{Meech11}).

\subsection{UV/visible emission features}

Photodissociation of water produces a lot of hydrogen in cometary comae, both directly and via further dissociation of OH, and the Lyman-$\alpha$ transition is one of the strongest emission features. However, it can only be observed from space, and even in Earth-orbit is difficult to detect in comets due to the overwhelming background from geocoronal emission. For the exceptionally bright and nearby comet C/1996 B2 (Hyakutake) Hubble Space Telescope (HST) spectroscopy detected the Ly-$\alpha$ emission, as the relative velocity (54 km s$^{-1}$) of the comet was sufficient to Doppler shift it away from the background \citep{combi1998}. Observations with small field-of-view (FOV), like HST, are difficult to interpret directly because the innermost coma is optically thick to Lyman-$\alpha$ \citep{combi1998}, so most observations have been very wide field like sounding rockets and the SOHO SWAN all-sky Lyman-$\alpha$ camera \citep{bertaux1998,makinen2001,combi2011}. Because the H coma can cover 10 or more degrees of the sky, background stars can become a serious limitation especially for fainter comets.  The lowest water production rates determined from SOHO SWAN measurements have been $\sim 10^{27}$ molec.\,s$^{-1}$, and these were only for comets very close to SOHO, e.g., comet 103P \citep{Combi2011-103P}, which was at a heliocentric distance of only slightly more than 1 au.  In table \ref{tab:comet-detections} we give an example from a Ly-$\alpha$ imager on another spacecraft, the 2$^\circ$ FOV LAICA camera on the experimental Japanese PROCYON micro-satellite, which was designed to test deep-space navigation of a 50 kg satellite. The LAICA camera was designed to study the geocorona, but was also successfully employed to measure water production rates in 67P near perihelion \citep{Shinnaka2017}. The small sizes of these telescopes (2.7 and 4.2 cm diameter for SWAN and LAICA, respectively) implies that more sensitive instruments could be built relatively easily, but background noise will remain a limiting factor for Ly-$\alpha$ observations. 

There are also emission lines from atomic oxygen in the UV, at 1304 and 1356 \AA, which we list in table \ref{tab:comet-detections} for completeness, but note that these are weak and haven't been used to derive production rates. Also only observable from space, they were detected in C/1995 O1 (Hale-Bopp) from sounding rocket observations \citep{McPhate1999}, and from close range in 67P by Rosetta/ALICE \citep{Feldman2015}, but the excitation process (photons vs electrons), and therefore the production rate, is highly model dependent (see section \ref{sec:Rosetta}). There is no realistic prospect of remote detection of these lines in a MBC.

Emission bands of OH, the direct product of water photo-dissociation, can be detected at optical wavelengths. The strongest group of OH lines visible in this wavelength range is in the 3070--3105 \AA\ region \citep{swings41}. This OH emission is often used to estimate the water production rate of  comets, via spectroscopy or photometry. Low resolution spectroscopy gives total production rates \citep[e.g][]{Cochran2012}, while higher resolution with either large telescopes or from space can improve signal-to-noise (S/N) for bright comets and reveal the structure within the band \citep[e.g][]{weaver2003,jehin2006}. Using narrow-band filters that isolate the region around the OH emission can give production rates and can be used to image the 2D structure of the gas coma, with OH production rates as low as $2.2 \times 10^{27}$ molec.\,s$^{-1}$  detected for comet 88P/Howell at 2 au from the Sun using the 60 cm TRAPPIST telescope (Opitom et al. 2017, in prep.), but the most sensitive production rates come from photometry with a traditional photomultiplier \citep[e.g.,][]{Schleicher2011}. However, at these wavelengths atmospheric extinction is high and the efficiency of most optical telescope detectors and optics is low, so OH emission is difficult to detect for faint comets. One attempt at detecting OH emission from a MBC was made, using the medium-resolution X-shooter spectrograph on the 8 m VLT, but only upper limits were obtained, at $8-9\times10^{25}$ molec.\,s$^{-1}$, in 2.5 hours of integration \citep{snodgrass-T1}. A search for OH emission from Ceres using VLT/UVES produced a similar upper limit \citep{Rousselot2011}, which is below the production rate found by Herschel observations \citep{Kueppers-ceres}, suggesting some variability in the outgassing rate from Ceres. 

\begin{landscape}
\begin{table}
\caption{Examples of detection of water and other key species in cometary comae.}
\label{tab:comet-detections}
\begin{tabular}{llllllllll}
\hline
Wavelength   & Species, line & Comet & Tel. / inst. & \O{} & SNR/s & $Q$  & $r$ & $\Delta$  & Ref. \\ 
&    &   &   & (m) & & (molec.\,s$^{-1}$) & (au) & (au) &  \\ 
\hline
1216 \AA & H, Ly-$\alpha$ & 67P/C-G & PROCYON/LAICA & 0.0415 & 7.07 & 1.24E+27 & 1.298 & 1.836 & 1  \\  
1304 \AA & [OI] &  note (a)\\ 
1356 \AA & [OI] &  note (a)\\
3080 \AA & OH & 73P-R/S-W 3 R & Lowell 1.1m/Phot. & 1.1 & 0.46 & 1.30E+25 & 1.029 & 0.074 & 2  \\ 
3080 \AA & OH & 88P/Howell & TRAPPIST & 0.6 & 0.14 & 2.20E+27 & 2 & 1.44 &  3  \\ 
5577 \AA & [OI] & 67P/C-G & VLT/UVES & 8.2 & 0.07 & 5.00E+27 & 1.36 & 1.94 & 4   \\ 
6300 \AA & [OI] & 67P/C-G & VLT/UVES & 8.2 & 0.47 & 5.00E+27 & 1.36 & 1.94 & 4   \\ 
6364 \AA & [OI] & 67P/C-G & VLT/UVES & 8.2 & 1.41 & 5.00E+27 & 1.36 & 1.94 & 4  \\ 
6563 \AA & H$\alpha$ & C/1996 B2 (Hyakutake) & McDonald & 2.7 & 0.24 & 3.00E+29 & 0.9 & 0.2 & 5  \\   
2.66 \micron & H$_2$O, $\nu_3$ & 22P/Kopff & Akari / IRC & 0.69 & 2.48 & 1.20E+27 & 2.4 & 2.4 & 6 \\   
2.9 \micron & H$_2$O, hot bands & C/2014 Q2 (Lovejoy) &	Keck/NIRSPEC & 10 & 0.75 & 5.90E+29 & 1.29 & 0.83 & 7 \\   
6.3 \micron & H$_2$O, $\nu_2$ & C/2004 B1 (LINEAR) & Spitzer / IRS & 0.85 & 3.17 & 1.00E+28 & 2.2 & 2.0 & 8 \\   
1665 GHz & H$_2$O, 212-101 & C/1995 O1 (Hale-Bopp) & ISO / LWS & 0.60 & 0.17* & 3.30E+29 & 2.8 & 3.0 & 9   \\ 
1113 GHz & H$_2$O & 81P/Wild 2 & Herschel/HIFI & 3.5 & 2.03 & 8.60E+27 & 1.61 & 0.93 & 10  \\   
557 GHz & H$_2$O & 81P/Wild 2 & Herschel/HIFI & 3.5 & 4.75 & 1.13E+28 & 1.61 & 0.93 & 10 \\   
183 GHz & H$_2$O & 103P/Hartley 2 & IRAM & 30 & 0.10 & 1.70E+25 & 1.06 & 0.15 & 11 \\   
18 cm & OH & 103P/Hartley 2 & Nan\c{c}ay & 100 & 0.10 & 1.70E+28 & 1.1 & 0.15 & 12  \\ 
\hline
1400-1640 \AA & CO, 4th+ (A-X) & C/2000 WM1 (LINEAR) & HST/STIS & 2.4 & 0.25 & 3.56E+26 & 1.084 & 0.358 & 13 \\ 
1850-2300 \AA & CO, Cameron & 103P/Hartley 2 & HST/FOS & 2.4 & 0.06 & 2.60E+27 & 0.96 & 0.92 & 14 \\ 
3883 \AA & CN & 67P/C-G & VLT/FORS2 & 8.2 & 0.12 & 1.40E+24 & 2.9 & 2.2 & 15  \\ 
3883 \AA & CN & C/2012 S1 (ISON) & Lowell 1.1m/Phot. & 1.1 & 0.07 & 1.29E+24 & 4.554 & 4.039 & 16  \\ 
4.26 \micron & CO$_2$, $\nu_3$ & 22P/Kopff & Akari / IRC & 0.69 & 2.32 & 1.30E+26 & 2.4 & 2.4 & 6 \\   
4.67 \micron & CO, $0\rightarrow 1$ & 29P/S-W 1 & VLT / CRIRES & 8.2 & 0.09 & 2.64E+28 & 6.3 & 5.5 & 17 \\   
265 GHz & HCN & 10P/Tempel 2 & JCMT & 15 & 0.08 & 9.00E+24 & 1.5 & 0.98 & 18 \\   
\hline
\end{tabular}
\\
* = estimated, no SNR or exposure time quoted in reference.\\
(a) [OI] lines in the UV were detected in C/1995 O1 (Hale-Bopp) via sounding rocket observations \citep{McPhate1999} and in 67P via Rosetta/Alice \citep{Feldman2015},
but production rates are not given as the parent(s) and excitation process is model dependent (see section \ref{sec:Rosetta}).\\
References: 1 = \citet{Shinnaka2017}; 2 = \citet{Schleicher2011} ; 3 = Opitom et al. 2017, in prep.; 4 = \citet{Jehin2015}; 5 = \citet{Combi1999}; 6 = \citet{Ootsubo2012}; 7 = \citet{Paganini2017}; 8 = \citet{Bockelee-Morvan2009}; 9 = \citet{Crovisier1997}; 10 = \citet{deVal-Borro2010}; 11 = \citet{Drahus2012}; 12 = \citet{Crovisier2013}; 13 = \citet{Lupu2007}; 14 = \citet{Weaver1994}; 15 = \citet{Opitom2017}; 16 = \citet{Knight2015}; 17 = \citet{Paganini2013}; 18 = \citet{Biver2012}.
\end{table}
\end{landscape}

Atomic oxygen emissions are detected in the optical range through three forbidden oxygen lines at 5577.339 \AA\ for the green line and 6300.304 and 6363.776 \AA\ for the red doublet. Those lines are produced by prompt emission following the photo-dissociation of various parent molecules ($\mathrm{H_2O}$, $\mathrm{CO_2}$, CO, but also $\mathrm{O_2}$) into a short-lived excited oxygen atoms \citep{festou81, cessateur16}.  The combination of measurements of two forbidden lines O$^1$D and O$^1$S shows that is possible to even estimate the CO$_2$/H$_2$O ratio \citep{mckay12,decock13,cessateur16}. The 6300.304 \AA\ line, which is the brightest of the three forbidden oxygen lines has been successfully used to derive water production rates \citep[e.g.,][]{spinrad82,mckay12}. Observation of the forbidden oxygen lines require high-resolution spectroscopy and a sufficient Doppler shift of the comet emission lines to be distinguished from telluric oxygen lines. Those observations are  challenging and cannot be done for all comets. However, they may be a very sensitive way to detect water (or CO and $\mathrm{CO_2}$) on faint comets, as observations of the  Rosetta target 67P showed that those oxygen lines were among the first emissions to be detected from Earth \citep{Jehin2015}. Such observations have been attempted on the MBC 133P with the UVES instrument at the VLT, but none of the forbidden oxygen lines were detected (D. Bodewits, Priv. Comm., 2017). Observations of Ceres and Themis also produced only upper limits of $4.6 \times 10^{28}$ and $4.5 \times 10^{27}$ molec.\,s$^{-1}$, respectively \citep{mckay17}.

Finally, for the UV/visible range, there is the emission line of H$\alpha$ at 6563~\AA. Despite its common use in astronomical observations, it has been detected in relatively few comets, and the best example of which we are aware was for Hyakutake, a very bright comet with $Q = 3 \times 10^{29}$ molec.\,s$^{-1}$, in spectroscopy with the 2.7 m telescope at McDonald observatory \citep{Combi1999}. H$\alpha$ does not appear promising for detection of outgassing from a MBC.

\subsection{Infrared and sub-mm/radio emission features}

\begin{figure}
\includegraphics[width=\columnwidth]{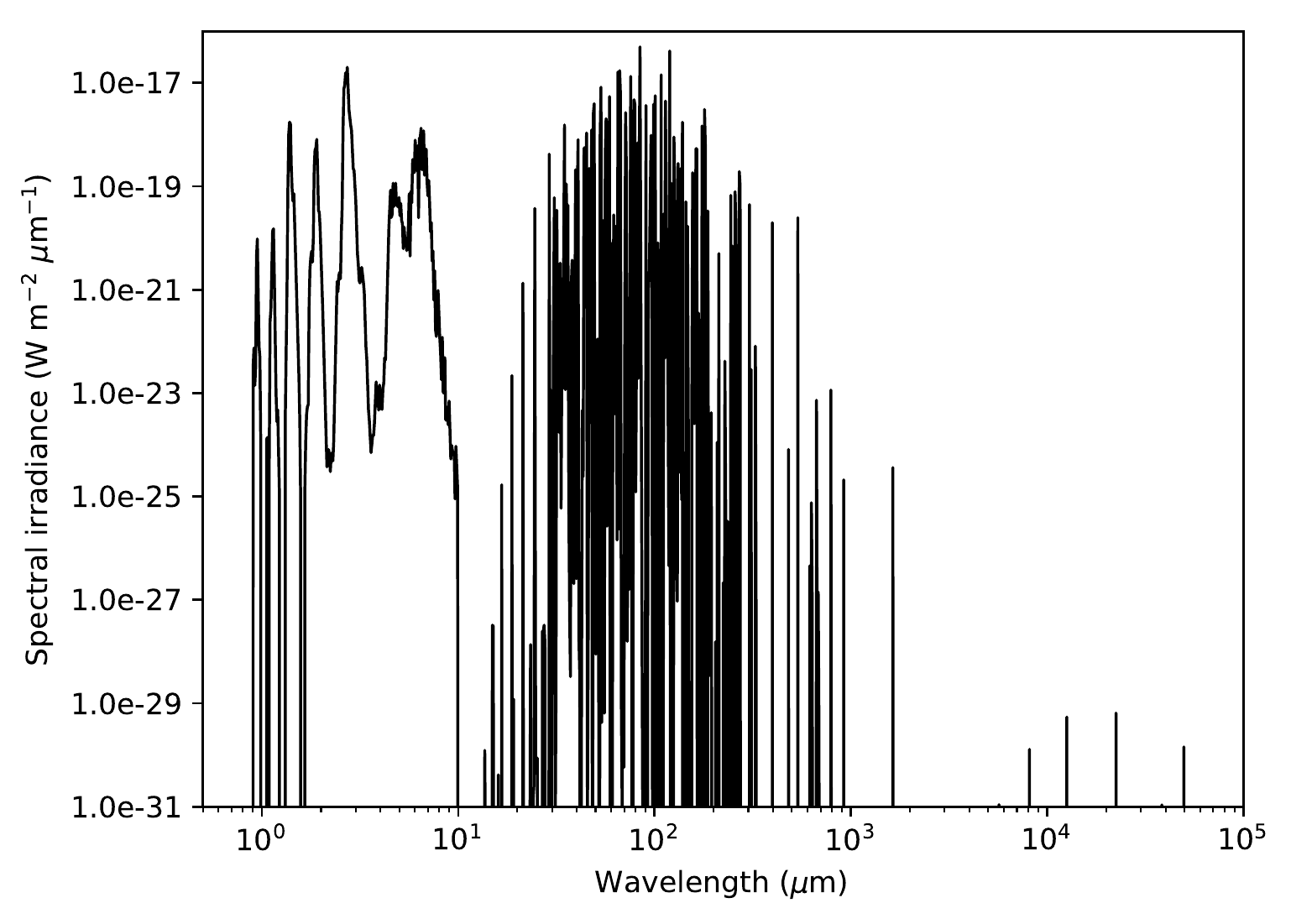}
\caption{Spectrum of water emission features in the NIR -- radio (1 $\mu$m -- 10 cm) at $R$=100, generated for a MBC with $Q = 10^{24}$ molec.\,s$^{-1}$, $r$ = 3 au, $\Delta$ = 2 au, and an atmospheric temperature of 80 K. The beam width is $1^\circ$ at all wavelengths.}
\label{fig:IRspec}
\end{figure}

Molecular species in the gas comae of comets, including water, produce a large number of emission features
 in the infrared \citep{crovisier:1984, mumma:2003,bockele-morvan:2004, cernicharo:2005} and sub-millimetre \citep{biver:2007, bockelee-morvan:2008},
 which are caused by the vibrational and rotational modes. We show the NIR--radio spectrum of a model MBC atmosphere comprised of water, generated using the Goddard Planetary Spectrum Generator\footnote{\url{https://ssed.gsfc.nasa.gov/psg/index.php}} \citep{PSG}, in fig.~\ref{fig:IRspec}. Water itself is difficult to observe from Earth due to its presence in the atmosphere, but at high resolution and with sufficient Doppler shift comet lines can be separated from telluric ones. Sufficiently high spectral resolution also allows the structure of vibrational bands to be resolved, or even isotopic measurements to be made for bright comets \citep[e.g.,][]{Paganini2017}. The advantage of observing in this region is that water is directly detected, rather than production rates being derived via daughter species. For fainter comets, where high-resolution spectroscopy is impossible, or where there is too low radial velocity with respect to Earth, observations from space telescopes are necessary. Some of the most sensitive observations were made with the Akari satellite, with detections of the 2.7 $\mu$m $\nu_3$ band of water at $Q = 1.2 \times 10^{27}$ molec.\,s$^{-1}$ made in comet 22P/Kopff when it was at 2.4 au from the Sun \citep{Ootsubo2012}. This band was also observed in bright comets in both imaging and spectroscopy by the International Space Observatory (ISO) satellite \citep{Colangeli1999,Crovisier1997} and the Deep Impact spacecraft \citep{Feaga2014}. At longer wavelengths, the $\nu_2$ band at 6.3 $\mu$m was observed by ISO and Spitzer, again for relatively bright comets with $Q \sim 10^{28} - 10^{29}$ molec.\,s$^{-1}$ \citep{Crovisier1997,Bockelee-Morvan2009}. None of these space telescopes are still operating at these wavelengths, but the James Webb Space Telescope (JWST) will (see section \ref{sec:future}).

At sub-mm and radio wavelengths very high resolution spectroscopy is possible, enabling detection of spectrally resolved individual rotational lines used to determine the kinematics of gas flows and excitation conditions in the coma, although mostly these observations are sensitive to larger molecules. With large dishes or arrays on Earth a large number of molecules have been identified in comets \citep[e.g.,][]{Biver2002,Biver2014}, especially hydrocarbons, with methanol and HCN being most regularly observed, and abundances normally measured relative to HCN\@. Water is generally not identified from the ground, although \citet{Drahus2012} were able to detect the 183 GHz band with the 30 m IRAM telescope during the close approach of comet 103P, at $Q=1.7\times 10^{25}$ molec.\,s$^{-1}$. The launch of sub-mm space telescopes, in particular Odin and \herschel{}, allowed observations of several ortho- and para-water  lines \citep[e.g., at 557 and 1113 GHz;][]{deVal-Borro2010, 2014A&A...564A.124D,hartogh2010,Biver2012}. The 3.5 m mirror of \herschel{} gave it excellent sensitivity, and the high resolution Heterodyne Instrument for the Far Infrared \citep[HIFI;][]{2010A&A...518L...6D} instrument meant that D/H could be measured for the first time in a bright enough JFC \citep[103P;][]{Hartogh2011}.

Direct detection of \ce{H2O} emission from MBCs has been attempted using \herschel{}.
Based on the visibility and anticipated gas emission activity during the \herschel{} mission lifetime (2009--2013), two MBCs, 176P and P/2012 T1, were observed by HIFI \citep{devalborro12,orourke13}.  176P passed its perihelion on 2011 June 30 and was observed by \herschel{}/HIFI on UT 2011 August 8.78 when it was at a heliocentric distance of \SI{2.58}{au} and a distance of \SI{2.55}{au} from the spacecraft.  At the end of the mission, the newly discovered P/2012 T1 was observed as part of director's discretionary time on UT 2013 January 16.31, about three months after its perihelion passage, when the object was at a heliocentric distance of \SI{2.50}{au} and a distance of \SI{2.06}{au} from \herschel{}.  The line emission of the fundamental \trans{} rotational transition of ortho-\ce{H2O} at \SI{557}{\GHz} was searched for in both objects.  
There was no detection of \ce{H2O} line emission in either of the targets. However, sensitive 3-$\sigma$ upper limits were inferred for the \ce{H2O} production rate of \SIlist{<4e25;<7.6e25}{\mols} for 176P and P/2012 T1, respectively.  While 176P was shown to be inactive during its 2011 passage from ground-based observations, dust emission activity was clearly observed in P/2012 T1 at the time of the \herschel{} observation suggesting that the gas production was lower than the derived upper limit \citep{orourke13}.  

Finally, at very long (radio) wavelengths, there is OH emission at 18 cm. There are two lines (1.665 and 1.667 GHz) that are regularly observed in comets, especially using the Nan\c{c}ay array \citep{Crovisier2013}. The (at the time) largest single dish telescope in the world, the 300 m Arecibo radio telescope, has also targeted comets at 18 cm \citep{Lovell2002}, but even using these facilities only the brightest comets are detectable, with the most sensitive detection again coming from the close approach of 103P to Earth in 2010. Detection of MBC-level outgassing at 18 cm does not appear likely.

\subsection{Absorption features from surface ice or coma ice grains}

It is also possible to detect water ice directly on surfaces of small bodies, or on grains in the comae of comets, through absorption features in the reflected solar continuum.
Water ice may take many forms, depending on the temperature and pressure at the time of formation \citep{petrenko:1999}. If temperatures were low (T $<$ 50K), water would have frozen out into its amorphous phase, but had they been higher (120--180~K), water molecules would have been able to arrange themselves into a crystalline structure \citep{Mukai1986}. Observationally, amorphous ice and crystalline ice are recognisable through infrared spectroscopy. The position, shape, intensity and width of absorption bands in an ice spectrum are indicators of the structure, temperature and thermal history of the ice \citep{newman:2008}. For ground-based observations, there are two key bands for recognising crystalline (as opposed to amorphous) water ice. One is the 3.1 $\mu$m Fresnel reflection peak, which is caused by a stretching mode of the water molecule, and another one is a temperature sensitive absorption band at 1.65 $\mu$m that is only seen in crystalline ice \citep{grundy:1998}. The 1.65 $\mu$m feature is absent in most available NIR spectra of the observed comets, except for the outbursting comet P/2010 H2 \citep{Yang:2010}. On the other hand, several thermodynamical models have proposed that the heat, generated through crystallisation of amorphous ice, is an important energy source thought to trigger distant cometary activity even when comets are beyond the critical sublimation heliocentric distance \citep{prialnik:1992,prialnik:2004}. Given the current observational constraints on water ice in comets, it remains uncertain whether the ice in comets is amorphous even before their entry to the hot, inner region of the Solar System. 

Most direct evidence of the presence of water-bearing minerals on asteroids comes from infrared observations, particularly in the so-called 3 $\mu$m region, where the hydroxyl fundamental absorption and the strong first overtone of water are both present. Characteristics of absorption features in this region, such as wavelength position of the band centers, the shape of the absorption bands and their relative intensities are diagnostics of mineralogy as well as abundance. However, the 3 $\mu$m region is notoriously difficult for ground-based observations because of intense atmospheric absorptions and faint solar radiation in this wavelength regime. Nevertheless, spectral surveys of large asteroids in the 3 $\mu$m region have flowered, mostly using the SpeX instrument at the IRTF telescope \citep{takir12,rivkin:2015}.

Compared with the NIR observations, visible spectroscopy is relatively easy to obtain using small to moderate-sized telescopes. Therefore, numerous efforts have been made to search for diagnostic features in MB asteroids between 0.4 -- 0.9 $\mu$m \citep{vilas:1985,luu:1990,sawyer:1991, xu:1994,bus:2002}. A shallow but fairly broad (width $\sim$ 0.25 $\mu$m) absorption feature centered near 0.7 $\mu$m is found to be very common among the low-albedo asteroids \citep{sawyer:1991}. This feature is attributed to an Fe$^{2+}$ $\rightarrow$ Fe$^{3+}$ charge transfer transition in hydrated minerals \citep{vilas:1989}. Consistently, a 0.7 $\mu$m feature is also seen in the laboratory spectra of CM2 carbonaceous chondrite meteorites and terrestrial phyllosilicates (a hydrated mineral), which is similar to the asteroidal feature both in band center and strength \citep{vilas:1989}.  Although \cite{king:1997} warned that a feature near 0.7 $\mu$m has been seen in many different minerals, and so this feature alone is not sufficient for identifying aqueous alteration, the majority of minerals that show an absorption centered at 0.7 $\mu$m are iron- and OH-bearing silicates \citep{rivkin:2002}. 

Additionally, thirteen C-, P-, and G-class asteroids were observed in the UV/blue spectral regions and they exhibit an absorption feature near 0.43 $\mu$m \citep{vilas:1993, cochran:1997}. The blue/UV drop-off observed in these low albedo asteroids is thought to be attributed to a ferric spin-forbidden absorption in aqueously altered iron-containing minerals \citep{burns:1981}.  Among low albedo asteroids, the UV drop-off is found to be correlated with the 3 $\mu$m absorption due to hydrated mineral components \citep{feierberg:1985}. Similarly, the same correlation is observed in carbonaceous chondrites, which contain a significant fraction of hydrated silicates, such as phyllosilicates. Therefore, it has been suggested that U-B index might be also useful as an indicator of the presence hydrated materials \citep{gaffey:1978}.  These UV/visible wavelength features have not yet been observed in MBCs.

\subsection{{\it in situ} detection of water}

The last set of techniques for detecting water are those that can be applied by visiting the target with a spacecraft. Various cometary missions have done so, with the recent Rosetta mission combining mass spectroscopy with remote sensing instruments that were sensitive to water via the various emission and absorption features described in previous sections. The mass spectrometer and pressure gauge instrument ROSINA on Rosetta first detected water in August 2014, at a distance of $\sim$ 100 km from the comet and 3.5 au from the Sun, at $Q \approx 5 \times 10^{25}$ molec.\,s$^{-1}$ \citep{Hansen2016}, although earlier detection was possible from the remote sensing instruments (see section \ref{sec:Rosetta}). Mass spectroscopy can measure both precise relative abundances of water and other volatiles, and isotopic ratios \citep{altwegg15}.

Suitably equipped spacecraft can also search for evidence of buried ice, in addition to identifying surface ice or outgassing water. Radar detection of subsurface water relies on the marked difference in dielectric properties between liquid water and other common geo-materials such as rocks and water ice. Whereas the relative dielectric permittivity of ice is 3.1, and that of rocks ranges between 4 and 10, liquid water has a permittivity of 80 \citep[][Appendix E]{ulaby1986microwave}. The marked dielectric contrast between a layer of liquid water and the surrounding rocks results in a high radar reflection coefficient, and thus in a strong radar echo. On Earth, this feature is routinely used in the identification of subglacial lakes \citep[see e.g.,][]{2013GeoRL..40.6154P}. While liquid water is not expected to be present on a MBC, radar would be useful in identifying rock and ice layers. The radar reflection coefficient of a surface is determined both by dielectric properties and by geometry. Rough surfaces diffuse the impinging radar pulse and weaken the backscatter towards the radar \citep{ogilvy}, thus making the identification of water more and more ambiguous as the water/rock interface becomes rougher or more curved.

The depth at which radar can detect echoes from a subsurface interface is affected by the properties of the material between the surface and the interface. Electromagnetic waves are absorbed in a dielectric medium, such as rocks and ices, and can be scattered by cracks, voids and other irregularities at scales comparable to the wavelength. However, assuming the results of the CONSERT experiment on board Rosetta can be used to predict the expected bulk dielectric properties of MBCs, then both dielectric attenuation and volume scattering should be low \citep{2015Sci...349b0639K}.

At this moment, there are no useful precedents for the detection of water within a MBC. The CONSERT experiment is based on the transmission of a radar signal through the cometary material, rather than on its reflection at a dielectric interface. If used on a MBC, CONSERT would probably be unable to identify a  ice layer, although it would likely measure a significant signal attenuation. The measurements at Phobos of the MARSIS experiment on board Mars Express \citep{2005Sci...310.1925P} are of limited value in predicting the expected performance of radar in detecting water within a MBC. MARSIS was originally designed solely for the observation of Mars, and Phobos is smaller than the MARSIS footprint \citep{2009epsc.conf..717S}.

A higher-frequency radar, such as SHARAD on board NASA's Mars Reconnaissance Orbiter \citep{2007Sci...317.1715S}, could be used to acquire full coverage of a MBC and produce its three-dimensional radar tomography. This is a well-known technique on Earth \citep[see e.g.,][]{knaell.cardillo}, but the irregular shape of a MBC will require some developments. In spite of the difficulties in deconvolving of the effects of geometric and dielectric interface properties on the echo, the detection of water pockets could then be demonstrated based on their greater radar backscatter cross section, compared to similar structures elsewhere in the subsurface, their relation to the surrounding stratigraphy, and the general geologic context.

\section{Expected activity levels for MBCs}
\label{sec:MBC-expected-levels}

When considering how best to detect evidence of sublimation from MBCs, it is important to consider what levels of activity can be reasonably expected, as well as which MBCs are most likely to exhibit detectable levels of activity.

\subsection{Predictions from energy balance}
\label{sec:energybalance}

As described in \citet{hsieh15_ps1} and elsewhere, the equilibrium temperature and unit-area water sublimation rate for a sublimating grey-body at a given heliocentric distance can be computed iteratively from the energy balance equation for a sublimating grey-body  (neglecting heat conduction),
\begin{equation}
\frac{F_{\odot}}{r^2}(1-A) = \chi\left[{\varepsilon\sigma T_{eq}^4 + L f_D\dot m_{w}(T)}\right],
\label{equation:sublim1}
\end{equation}
the sublimation rate of ice into a vacuum,
\begin{equation}
\dot m_{w} = P_v(T) \sqrt{\frac{\mu}{2\pi k T}},
\label{equation:sublim3}
\end{equation}
and the Clausius-Clapeyron relation,
\begin{equation}
P_v(T) = 611 \times \exp\left[{\frac{\Delta H_{subl}}{ R_g}\left(\frac{1}{273.16} - \frac{1}{T}\right)}\right] .
\label{equation:sublim4}
\end{equation}
In the energy balance equation, $T_{eq}$ is the equilibrium surface temperature, $F_{\odot}=1360$~W~m$^{-2}$ is the solar constant, $r$ is the heliocentric distance of the object in au, $A=0.05$ is the assumed Bond albedo of the body, $\chi$ describes the distribution of solar heating over an object's surface ($\chi=1$ for a flat slab facing the Sun where this so-called subsolar approximation produces the maximum attainable temperature for an object, $\chi=\pi$ for the equator of a rapidly rotating body with zero axis tilt, and $\chi=4$ for an isothermal sphere, as in the limiting approximation of an extremely fast rotator and strong meridional heat flux), $\varepsilon=0.9$ is the assumed effective infrared emissivity, and $\sigma$ is the Stefan-Boltzmann constant.  In the sublimation rate equation, $L=2.83$~MJ~kg$^{-1}$ is the latent heat of sublimation of water ice, which is nearly independent of temperature, $f_D$ describes the reduction in sublimation efficiency caused by the diffusion barrier presented by a rubble mantle, where $f_D=1$ in the absence of a mantle, $\dot m_w(T)$ is the water mass loss rate due to sublimation of surface ice, $\mu=2.991{\times}10^{-26}$~kg is the mass of one water molecule, $k$ is the Boltzmann constant. The equivalent ice recession rate, $\dot \ell_{i}$, corresponding to $\dot m_{w}$ is given by
$\dot \ell_{i} = \dot m_{w}/ \rho$,
where $\rho$ is the bulk density of the object.  Lastly, in the Clausius-Clapeyron relation, $P_v(T)$ is the vapour pressure of water in Pa, $\Delta H_{subl}=51.06$~MJ~kmol$^{-1}$ is the heat of sublimation for ice to vapor, and $R_g=8314$~J~kmol$^{-1}$~K$^{-1}$ is the ideal gas constant.  

\setlength{\tabcolsep}{3pt}
\begin{table}
\caption{MBC activity properties 
}
\label{tab:mbcactivity}       
\begin{tabular}{lcccccccl}
\hline\noalign{\smallskip}
Object & $q$$^a$  &  $R_N$$^b$ & $v_{esc}$$^c$ & \multicolumn{2}{c}{$Q({\rm H_2O})_{q}$$^d$} & $dM/dt$$^e$ & Active Range$^f$ & Ref.$^g$ \\
\noalign{\smallskip}\hline\noalign{\smallskip}
     133P & 2.664 & 1.9  & 2.2 &  3.5$\times$10$^{16}$ & 1.6$\times$10$^{24}$ & 1.4 & 350$^{\circ}$--109$^{\circ}$  & [1-4]  \\ 
     238P & 2.366 & 0.4  & 0.5 &  1.1$\times$10$^{17}$ & 2.2$\times$10$^{23}$ & 0.2 & 306$^{\circ}$--123$^{\circ}$  & [5,6] \\ 
     176P & 2.580 & 2.0  & 2.4 &  4.9$\times$10$^{16}$ & 2.5$\times$10$^{24}$ & 0.1 & 1$^{\circ}$--19$^{\circ}$  & [1,7] \\ 
     259P & 1.794 & 0.3  & 0.4 &  5.0$\times$10$^{17}$ & 5.7$\times$10$^{23}$ & --- & 315$^{\circ}$--49$^{\circ}$  & [8-10] \\ 
     324P & 2.620 & 0.6  & 0.7 &  4.2$\times$10$^{16}$ & 1.9$\times$10$^{23}$ & 4.0 & 300$^{\circ}$--96$^{\circ}$  & [11-13] \\ 
     288P & 2.436 & 1.3  & 1.5 &  8.4$\times$10$^{16}$ & 1.8$\times$10$^{24}$ & 0.5 & 338$^{\circ}$--47$^{\circ}$  & [14-16] \\ 
P/2012 T1 & 2.411 & ---  & --- &  9.3$\times$10$^{16}$ & ---                  & $\sim$1 & 7$^{\circ}$--48$^{\circ}$  & [17-18] \\
P/2013 R3 & 2.204 & ---  & --- &  1.8$\times$10$^{17}$ & ---                  & --- & 14$^{\circ}$--43$^{\circ}$  & [19] \\
     313P & 2.391 & 0.5  & 1.2 &  1.0$\times$10$^{17}$ & 1.3$\times$10$^{24}$ & 0.4 & 354$^{\circ}$--53$^{\circ}$  & [20-23] \\ 
P/2015 X6 & 2.287 & ---  & --- &  1.4$\times$10$^{17}$ & ---                  & 1.6 & 329$^{\circ}$--344$^{\circ}$  & [24] \\
P/2016 J1-A & 2.448 & $<$0.9  & $<$1.1 &  8.1$\times$10$^{16}$ & ---                  & 0.7 & 346$^{\circ}$--12$^{\circ}$  & [25,26] \\
P/2016 J1-B & 2.448 & $<$0.4  & $<$0.5 &  8.1$\times$10$^{16}$ & ---                  & 0.5 & 346$^{\circ}$--12$^{\circ}$  & [25,26] \\
\noalign{\smallskip}\hline
\end{tabular}
\smallskip
\newline $^a$ Perihelion distance, in au.
\newline $^b$ Effective nucleus radius, in km.
\newline $^c$ Theoretical escape velocity, m~s$^{-1}$, in assuming spherical, non-rotating nuclei with bulk densities of $\rho\sim2500$~kg~m$^{-3}$.
\newline $^d$ Unit-area and total (assuming 100\% active area) theoretical water sublimation rates, in molec.~m$^{-2}$~s$^{-1}$ and molec.~s$^{-1}$, at perihelion, assuming a spherical sublimating isothermal gray-body. 
\newline $^e$ Peak reported observed dust mass production rate in kg s$^{-1}$.
\newline $^f$ True anomaly range over which activity has been reported. Listed ranges are incomplete, sometimes reflecting limitations in observational coverage rather than the confirmed absence of activity over orbit arcs not included in the listed true anomaly ranges.
\newline $^g$ References:
[1] \citet{hsieh09_alb}; [2] \citet{jewitt14_133p}; [3] \citet{hsieh10_133p}; [4] \citet{kaluna11_133p};
[5] \citet{hsieh09_238p}; [6] \citet{hsieh11_238p};
[7] \citet{hsieh11_176p};
[8] \citet{jewitt09}; [9] \citet{maclennan12}; [10] \citet{hsieh17_259p};
[11] \citet{moreno11_324p}; [12] \citet{hsieh12_324p}; [13] \citet{hsieh14_324p};
[14] \citet{hsieh12_288p}; [15] \citet{licandro13_288p}; [16] \citet{agarwal16_288p};
[17] \citet{hsieh13_p2012t1}; [18] \citet{moreno13_p2012t1};
[19] \citet{jewitt14_p2013r3};
[20] \citet{hsieh15_313p}; [21] \citet{jewitt15_313p}; [22] \citet{jewitt15_313p_2}; [23] \citet{pozuelos15_313p};
[24] \citet{moreno16_p2015x6};
[25] \citet{hui17_16J1}; [26] \citet{moreno17_16J1}
\end{table}

\begin{figure}
\includegraphics[width=1.0\columnwidth]{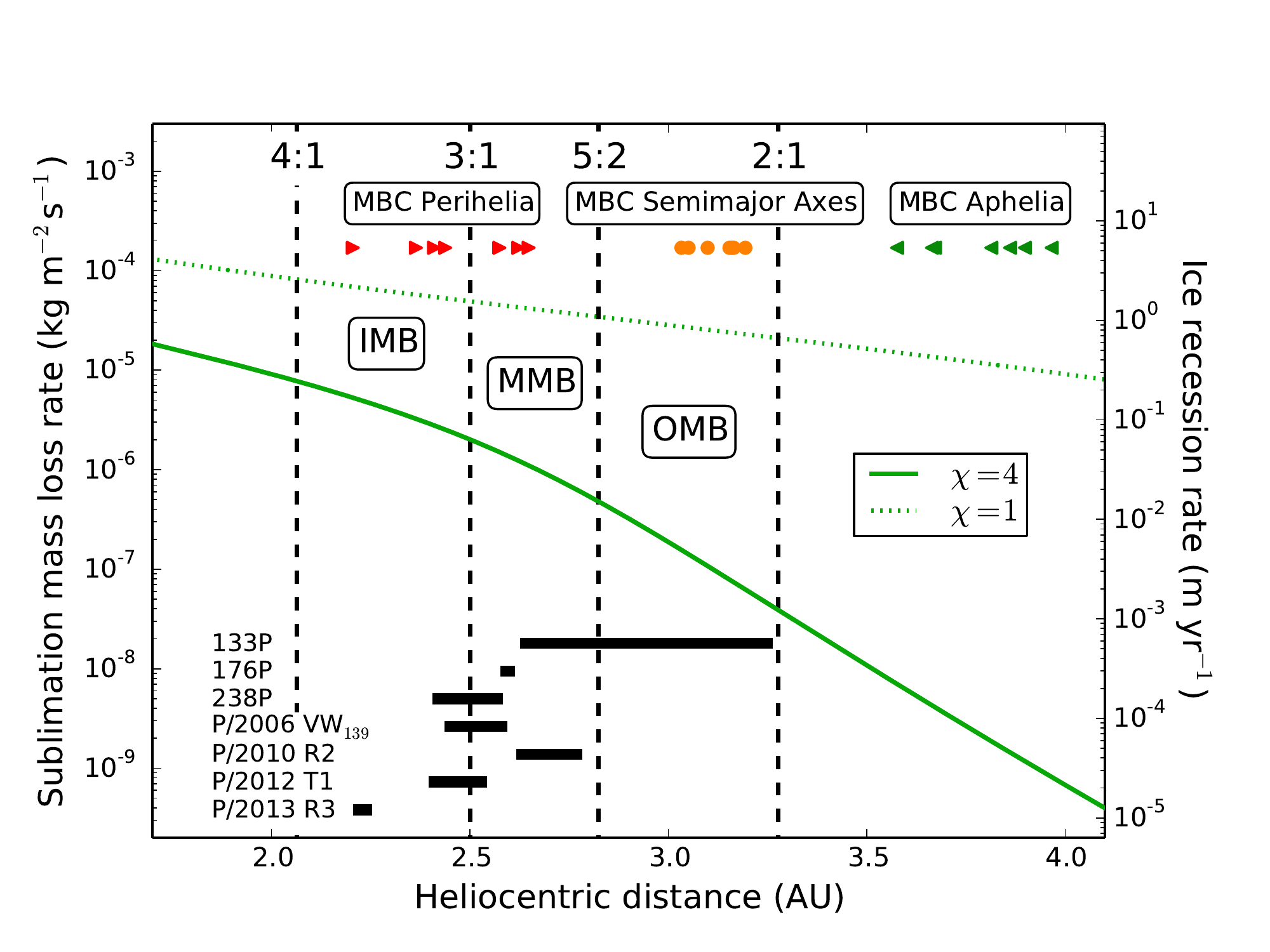}
\caption{Mass loss rate due to water ice sublimation from a sublimating grey body as a function of heliocentric distance over the range of the main asteroid belt, where the semimajor axis ranges of the inner, middle, and outer MB are labeled IMB, MMB, and OMB, respectively, and mass loss rates calculated using the isothermal approximation ($\chi=4$) and subsolar approximation ($\chi=1$) are marked with solid and dashed green lines, respectively.  The positions of the major mean-motion resonances with Jupiter (4:1, 3:1, 5:2, and 2:1) that delineate the various regions of the main asteroid belt are shown with vertical dashed black lines.  Also plotted are the perihelion distances (red, right-facing triangles), semimajor axis distances (orange circles), and aphelion distances (green, left-facing triangles) of the known outer main-belt MBCs, as well as the range of heliocentric distances over which they have been observed to exhibit activity (thick black horizontal lines). From \citet{hsieh15_ps1}.}
\label{fig:sublimrate}
\end{figure}

We use these equations to compute the peak expected sublimation rate using the isothermal approximation for the known MBCs, and also convert unit-area sublimation rates from the above equations to a total surface-wide sublimation rate for each object assuming spherical bodies and active areas of 100\% (Table~\ref{tab:mbcactivity}).  We can see from these results that there is a range of expected maximum water sublimation rates expected from various MBCs, and two important factors influencing these maximum rates are perihelion distance and nucleus size.  \citet{hsieh15_ps1} noted that, on average, the currently known MBCs have higher eccentricities (and therefore smaller perihelion distances) than the overall outer main-belt asteroid population, suggesting that the activity of those particular objects could be at least partly due to the fact that they reach higher peak temperatures than other asteroids with similar semimajor axes.  The significance of having larger-than-average eccentricities and therefore smaller perihelion distances is further illustrated by Figure~\ref{fig:sublimrate}, which illustrates water ice sublimation rates as a function of heliocentric distance, computed using Equations~\ref{equation:sublim1}, \ref{equation:sublim3}, and \ref{equation:sublim4} as well as the positions of the perihelion, aphelion, and semimajor axes of seven of the known MBCs.  It is interesting to note that over the ranges of heliocentric distances traversed by these MBCs over the course of their orbits, the theoretical water ice sublimation rate varies by as much as four orders of magnitude in the isothermal case. This suggests that the fact that all are seen to be active near perihelion is not just a coincidence or observational bias.

Meanwhile, we see that larger maximum total water sublimation rates are expected for larger objects than smaller objects due to the simple fact that larger objects have greater surface areas over which sublimation can occur.  This presents an interesting dilemma since larger (and therefore more massive) objects also have larger escape velocities (cf.\ Table~\ref{tab:mbcactivity}) that ejected dust grains must reach in order to overcome the gravity of the active body and escape into space where they can be observed as comet-like dust emission.  Ejection velocities for dust emission from many MBCs tend to be relatively low compared to ejection velocities for classical comets, yet they are often comparable to the escape velocities of the corresponding nuclei \citep[cf.][]{hsieh04_133p}. This competition between size preferences may be a highly consequential effect, meaning that there may be an `ideal' size or size range for detectable MBCs where they are large enough to produce observable amounts of gas or dust emission but small enough that ejected dust can actually escape into space to become observable in the first place.  \citet{hsieh09_htp} also noted that if MBC ice must be preserved in subsurface layers in order to survive to the present day and only excavated recently \citep[cf.][]{hsieh04_133p}, currently active MBCs must also tend to somewhat larger sizes so that they have the surface area to experience impacts capable of excavating subsurface ice on timescales consistent with their discovery rates in the asteroid belt, but still must be small enough for ejected dust to reach the escape velocity of the emitting body.

Interestingly, we see that measures of activity strength such as the peak inferred dust production rate or total activity duration listed in Table~\ref{tab:mbcactivity} are not strongly correlated to either the theoretical unit-area or total expected water sublimation rates at perihelion.  This discrepancy could be due to a number of factors, including differences in initial ice content or distribution, devolatilisation progression of active sites, overall fractions of active surface area on each body, or thermal properties of each object.  In particular, rotation rates are not known for most MBC nuclei, yet may be significant factors in whether it is most appropriate to model their thermal behavior using either the subsolar or isothermal approximations.

Notably, the total maximum predicted water sublimation rates listed in Table~\ref{tab:mbcactivity} are smaller than most of the detection limits for various sublimation detection techniques discussed elsewhere in this paper, and also assume 100\% active areas, which are almost certainly not the case for most or all MBCs  (for comets active areas of a few percent or less are typical; \citealt{ahearn95}).  We note, however, that these values represent calculations based on the isothermal (or `fast-rotator') approximation, which produces the lowest temperatures and lowest production rates.  At the other extreme, the sub-solar approximation gives production rates as much as two orders of magnitude larger, suggesting that we may be more likely to detect gas emission from slower-rotating MBC nuclei or those with higher thermal inertias.  Additionally, the total maximum predicted water sublimation rates listed here are computed assuming the surface area of spherical nuclei, whose effective radii have been estimated from their scattering cross-sections.  If MBC nuclei are irregularly shaped, like 67P, or perhaps even binary, like 288P \citep{agarwal16_288p}, larger usable surface areas and therefore higher gas production rates could be possible.

\subsection{Predictions from observed dust activity}
Although no direct detection of gas in a MBC has been made to date, we can use the observed dust comae as indicators of the mass-loss rates from these bodies. One method is to model the observed dust distribution to find the probable grain size distribution, the dust mass ejected over time, and then use nominal gas to dust mass loss ratios to constrain the sublimation rate. \cite{moreno11_324p} derived a maximum sustained dust mass loss rate of $\sim 3-4$ kg~s$^{-1}$ at 324P. In similar studies,  \citet{moreno13_p2012t1} found that P/2012 T1 emitted $6-25\times10^6$ kg of dust over 4--6 months, implying an average mass loss rate of $\sim 2$ kg~s$^{-1}$, while \citet{pozuelos15_313p} found dust mass loss rates of $\sim 0.2-0.8$ kg~s$^{-1}$ at 313P.  Assuming a dust/gas mass ratio of $\sim$1:1, this would imply maximum water sublimation rates of $Q({\rm H_2O})\sim10^{26}$ molec.\,s$^{-1}$. 

Another method is by direct comparison with  observations of normal active comets. The standard method of estimating the dust content of the coma is through measurement of the $Af\rho$ parameter (units of cm) as defined by \cite{ahearn1984}. Although transformation to  a dust mass-loss rate requires further assumptions or modelling, $Af\rho$ has the advantage that it is calculated solely on the basis of the observed photometric brightness. At the same time, the production rate $Q({\rm gas})$ can be accurately constrained through narrow-band photometry or spectrophotometry. \cite{ahearn95} reported photometric observations and analysis of tens of comets, finding that for a typical comet  $\log[Af\rho/Q({\rm OH})]=-25.8\pm0.4$. The most active MBCs discovered so far appear to have $Af\rho\simeq 10-20$ cm (P/2012 T1, \citealt{hsieh13_p2012t1}; 313P, \citealt{Hui2015}). Using the above relationship would imply $Q({\rm H_2O})\simeq(7^{+10}_{-4})\times10^{25}$ molec.\,s$^{-1}$ assuming $Q({\rm OH})\simeq 0.9\:Q({\rm H_2O})$.

Comparing these estimates of $Q({\rm H_2O})$  with the upper limits in table \ref{tab:CNlimits} implies that current observational efforts may be close to directly detecting gas in MBCs. But there are two important factors to take into account. First, many upper limits are based on the non-detection of the bright CN emission band and assuming normal CN/OH cometary ratios. As discussed in section 4.2, modelling indicates that more volatile species in subsurface ices may be depleted due to thermal processing. If HCN is the photodissociation parent molecule of CN, then the CN/OH ratio in MBC comae may be lower than in a normal comet.  This would imply that $Q({\rm H_2O})$ is higher than expected from the measured upper limit to  $Q({\rm CN})$. Second,
mantling of the source region could occur to due fallback of slowly moving dust grains
back onto the surface of the MBC. This `airfall' was readily apparent in images of 67P from the Rosetta spacecraft \citep{thomas2015}. This could plausibly increase the dust/gas ratio over time as the airfall layer on the sublimation site increases and the thermally active ice surface diminishes. This would mean that $Q({\rm H_2O})$ is less than anticipated from measurements of the dust coma.

\subsection{Relative strength of water signatures}

We now consider which of the various water detection methods listed in section \ref{sec:water-obs} is most promising to detect MBC-level outgassing, which, following the discussion in the previous subsections, we take as $Q$(H$_2$O) = $10^{24}$ molec.\,s$^{-1}$.
Comparing the relative effectiveness of the listed techniques is not straightforward, due to the great differences in observation type, geometry, and activity level of the comets used as examples. Even observations of the same comet -- C/2009 P1 (Garradd) -- with different telescopes/techniques gave different production rates, which was attributed to a halo of icy grains and different FOVs \citep{Combi2013}. Nevertheless, we attempt to draw some approximate conclusions by scaling these observations with a number of simplifying assumptions. We discard the possibility of {\it in situ} investigation for now; although it would certainly be  effective, it is a very different prospect to astronomical observation in terms of cost. We also leave detection of absorption features out of this comparison, as this appears to be only possible for larger bodies: even for much more active comets, surface ice features are not detected remotely, and even {\it in situ} exploration shows only relatively small and variable ice patches on surfaces \citep{DeSanctis2015}.

\begin{figure}
\includegraphics[width=0.8\columnwidth]{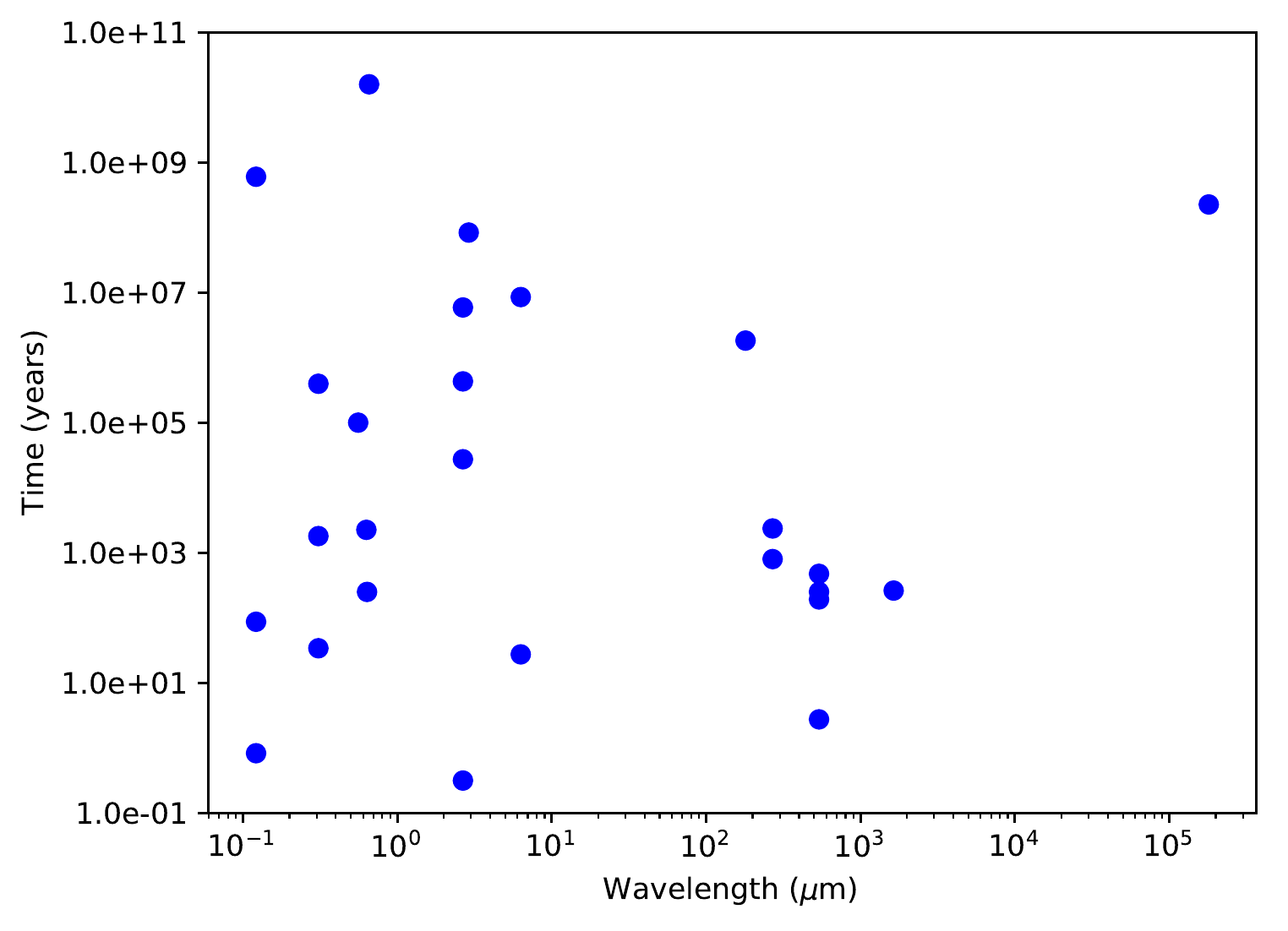}
\caption{Exposure time required to get 5$\sigma$ detection of $Q = 10^{24}$ molec.\,s$^{-1}$ outgassing from a MBC, based on scaling the comet detections at various wavelengths in table \ref{tab:comet-detections}.}
\label{fig:detect-MBC}
\end{figure}

To compare the strength of emission features across a range of wavelengths we scale the observations listed in table \ref{tab:comet-detections} to give the exposure time required to achieve a S/N of 5 detection of $Q = 10^{24}$ molec.\,s$^{-1}$  at $r$ = 3 au and $\Delta$ = 2 au. We ignore velocity effects, both heliocentric (i.e. Swings effect) and geocentric (i.e. assuming that lines are detectable without telluric interference). We make the assumption that the aperture is of fixed angular size and that the coma distribution scales as $\rho^{-1}$, i.e. that the signal scales as $\Delta^{-1}$, which is not necessarily the case. A better approximation could be made with scaled Haser models \citep{Haser1957} for each observation, taking into account the different slit or measurement aperture areas, but we judged this to be of secondary importance given the other uncertainties. We plot the scaled exposure times needed to make 5$\sigma$ detections as a function of wavelength in fig. \ref{fig:detect-MBC}. It is clear that the exposure times required with current technology are infeasible, and that there is a very large range  (from $\sim$1 year to a Hubble time!). There are minima at Ly-$\alpha$, the 2.66 $\mu$m water band, and at 557 GHz, indicating that these are the most promising places to try for future detections, although none of the space telescopes used to make the sensitive detections at these wavelengths (PROCYON/LAICA, Akari, Herschel) remain operational today. Of the ground-based observations, the most promising is the detection of the 3080 \AA{} OH band with the photometer at the Lowell 1.1-m.

\begin{figure}
\includegraphics[width=0.8\columnwidth]{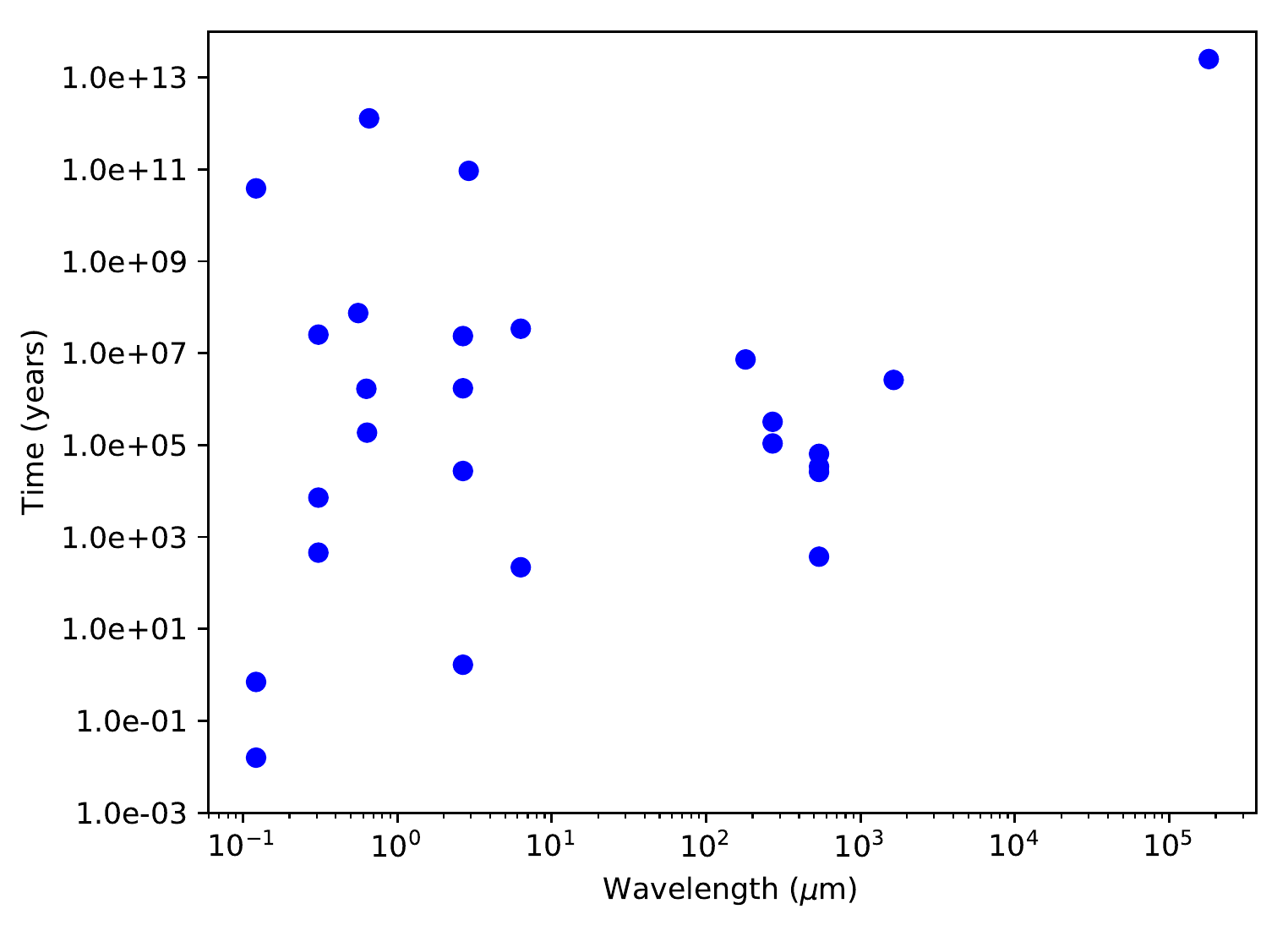}
\caption{Same as fig. \ref{fig:detect-MBC}, but scaled to a telescope diameter of 30 cm  in space.}
\label{fig:detect-MBC-scaled}
\end{figure}

To enable a slightly fairer comparison between the different telescopes, we further scaled the exposure times to the same diameter of telescope. This also involves the significant assumption that telescope sensitivity can simply be scaled by collecting area, and obviously ignores the fact that a 10 m radio dish is a much cheaper instrument than a 10 m optical telescope. For comparison, we chose to scale the observations to a 30 cm diameter telescope (fig. \ref{fig:detect-MBC-scaled}), as could be imagined in a relatively cheap space telescope in Earth orbit (e.g., the ESA S-class CHEOPS). Here the the strong Ly$\alpha$ bands are clear winners, which is not surprising given the small size of current telescopes at this wavelength, but this ignores the significant background issues in these observations -- S/N will not scale directly with telescope collecting area in this case. The NIR region continues to be promising, and a small-to-mid-sized space telescope covering 2.66 $\mu$m could be of use for a broad survey of weak outgassing. In the immediate future, the JWST will cover this range, but will not be used for broad surveys (see section \ref{sec:future}).


\section{Rosetta}\label{sec:Rosetta}

The recent ESA mission Rosetta has been transformative in cometary science, which includes implications for MBCs, at least by providing a very detailed source of information on a JFC for comparison. In this section we briefly review some of the more relevant findings, in particular considering the early phase of the mission, when Rosetta first encountered comet 67P as its activity was only just starting. We are still in the early days of analysis of Rosetta data, but a review of some of the key results to date is presented by \citet{Taylor-PTRSA}.

Rosetta entered orbit around 67P in August 2014, when the comet was at 3.6 au from the Sun, but was already able to study the beginning of activity while approaching using its remote sensing instruments, with the first detections of the comet and its dust in images taken as early as March 2014 \citep{Tubiana2015}. The first detection of the gas coma was through sub-mm observations of the 557 GHz water band by the MIRO instrument, in June 2014 at 3.9 au from the Sun, when the spacecraft was around half a million km from the comet \citep{Gulkis2015Sci}. At this time the water production rate was slightly below the strongest limits on MBCs to date, at $1 \times 10^{25}$ molec.\,s$^{-1}$, and outgassing was not detectable from Earth (via sensitive searches for CN with large telescopes -- \citealt{Snodgrass2016,Opitom2017}). This is the lowest activity comet environment ever visited by a spacecraft, with all previous mission performing fast flybys at $\sim$1 au, and there were some surprising results: The Rosetta Plasma Consortium instruments discovered oscillations in the magnetic field at around 40 mHz attributed to interactions with cometary ions and the solar wind \citep{Richter2015}, the so-called `singing comet' based on the public release of an audio version of this interaction. \citet{Feldman2015} show that emission lines from atomic hydrogen and oxygen, observed by the Alice instrument in the UV, could only be explained by electron impact dissociation rather than the more typical photodissociation seen in cometary comae. This effect was also necessary to explain the morphology of the gas in the inner coma, as imaged by the OSIRIS cameras \citep{bodewits16}. Both of these effects were no longer detectable closer to perihelion, when the comet had a more typical activity. 

\begin{figure}
\includegraphics[width=1.0\columnwidth]{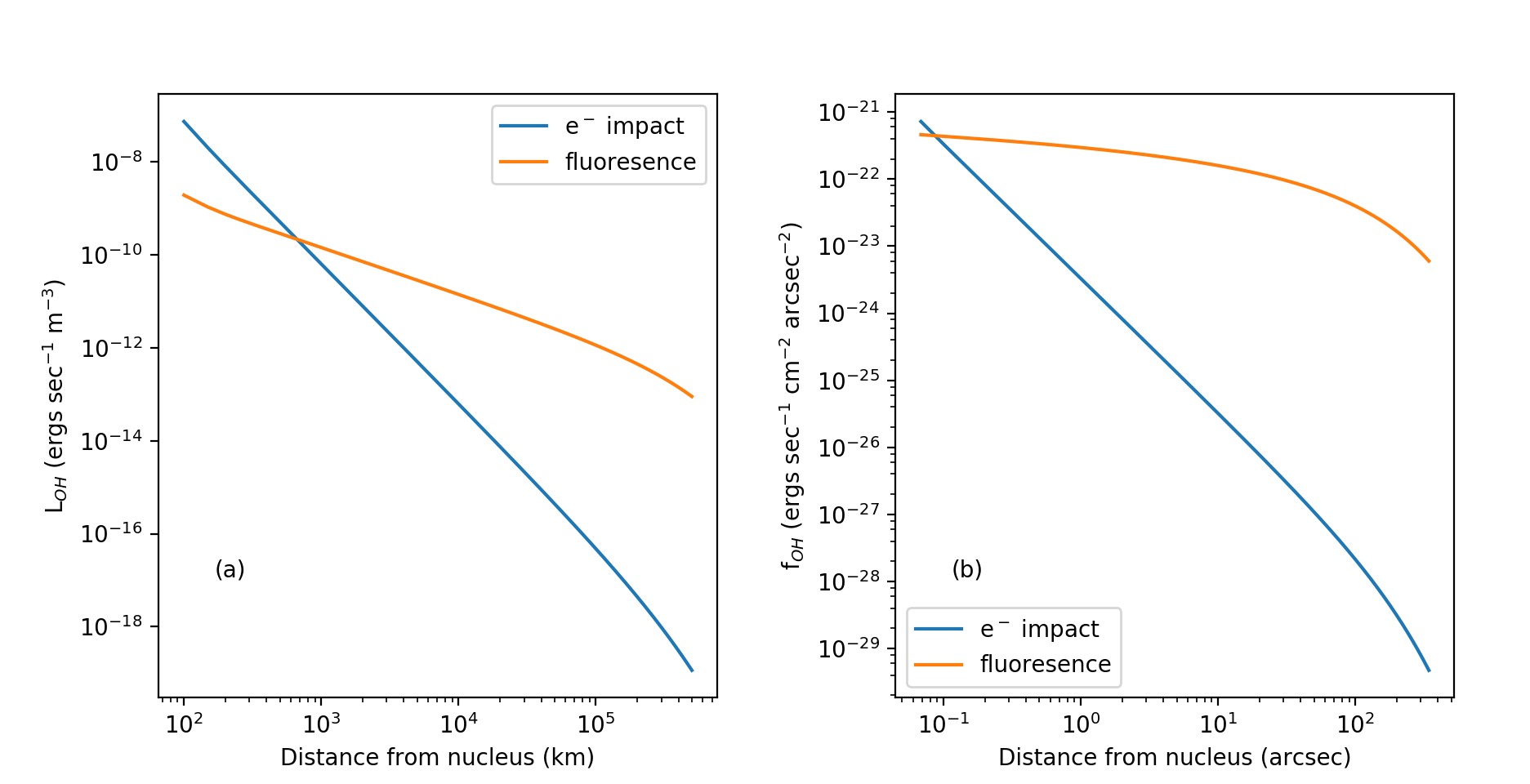}
\caption{(a) Calculated specific intensities of the OH (0-0) band at 3080~\AA\ due to both fluorescence and prompt emission following electron impact dissociation of water, as a function of radial distance from the nucleus. The underlying H$_2$O and OH densities were calculated assuming a Haser model with $Q({\rm H_2O})=\SI{e24}{\mols}$. (b) Integrated surface brightness profiles observed at Earth derived from this model, assuming the MBC is at $r=3$ au, $\Delta=2$ au.}
\label{fig:e-impact}
\end{figure}

A MBC with activity levels similar to, or even lower than, 67P at $\sim$3.5 au, could be expected to show similar physics. While testing its interaction with the solar wind will require {\it in situ} probing with a spacecraft equipped with a magnetometer, it is probable that electron impacts would affect the observed emission in gas lines. To estimate the magnitude of this effect, we assume that electron impact dissociation of H$_2$O into the $A^2\Sigma^+$ state of OH takes place, followed by prompt emission to the ground state $X^2 \Pi$. The increase in extra emission is extremely sensitive to the production rate of water, as the photoionisation of H$_2$O is expected to be the primary source of electrons in the inner coma. Scaling the conditions at 67P to a MBC at $\sim3.0$ au with a production rate of $Q({\rm H_2O})=10^{24}$ molec.\,s$^{-1}$, the specific
luminosity of the OH coma in ergs cm$^{-2}$ s$^{-1}$ is shown in fig.~\ref{fig:e-impact}. We model the density distribution using a simple Haser model, and scale the Rosetta measured electron density distribution to estimate the electron impact induced emission using the technique described in \cite{bodewits16}. A nearby 
spacecraft should be sensitive to this emission. However, integrating this luminosity distribution into potential surface brightnesses as seen from Earth shows that while a significant brightness enhancement exists, it is located within 1 arcsec of the MBC at low apparent flux levels.

Aside from exploring a very low activity comet environment, Rosetta is also important in the context of MBCs as it reveals a larger picture of comet formation. The full implications are likely to be debated for some time as further analysis of the Rosetta data is performed, but first attempts to model comet formation based on Rosetta results \citep[e.g.,][]{Davidsson2016} point to the low density / high porosity of the comet \citep{Sierks2015Sci} and the presence of hyper-volatile species such as O$_2$ and N$_2$ \citep{Bieler2015,Rubin2015} to show that it must have formed far from the Sun and avoided significant heating during its formation and subsequent evolution. This must be significantly different to the MBC case discussed in section \ref{sec:ice-survival}, and further points to an expected difference in the observed composition of outgassing species between MBCs and JFCs.


\section{Future prospects}\label{sec:future}

In this section we consider what options we will have for searching for water in MBCs in the coming years, both from ground- and space-based observatories, and from proposed dedicated space missions. 

\subsection{Ground-based telescopes}

It is worth considering if a traditional photoelectric photometer mounted on a large telescope might allow successful detection of gas from a MBC. A photometer performs like a single large pixel, allowing it to achieve a particular signal-to-noise more efficiently than a CCD, where the noise associated with bias and readout becomes significant due to the large number of pixels in an aperture. Assuming a system could be scaled up directly, a photometer on a 10 m class telescope would provide a factor of $\sim$100 improvement over existing facilities, e.g., the 1.1-m telescope at Lowell Observatory used by D.~Schleicher. \citet{Knight2015} detected CN in 30 minutes on comet C/2012 S1 ISON at $r=4.55$~au with a production rate of $Q$(CN) $\sim$ 10$^{24}$~molec.\,s$^{-1}$, near the limit for this instrument. This implies a hypothetical detection capability of 10$^{21}-10^{22}$~molec.\,s$^{-1}$ when accounting for the larger mirror, longer integrations, and a MBC at smaller $r$ and $\Delta$. Such production rates are near or somewhat below the most restrictive upper limits placed on MBC activity to date (as discussed in Section~5). Although the capabilities are promising, the lack of spatial context needed to definitively detect a very faint coma near a relatively bright nucleus likely make a photometer a sub-optimal choice.

 We briefly consider narrowband imaging with a CCD, which would have
advantages over a photometer in providing spatial context, having
generally higher quantum efficiency, and offering the possibility of
selecting arbitrarily sized apertures. However, despite the very low
read noise of modern CCDs, the small number of photons spread across a
very large number of pixels makes the detection of faint gas emission
via imaging relatively inefficient: with a blue optimized system and judicious binning, a CCD
on a 10 m may be able to go 1--2 orders of magnitude fainter than a
photometer on a 1 m with the same exposure times. However, this still
likely under-performs the hypothetical photometer on a 10 m by about an
order of magnitude.

With the start of scientific operations of the Atacama Large Millimetre/submillimetre Array (ALMA) interferometer in the second half of 2011, several observations of moderately bright Oort Cloud comets have been carried out with high sensitivity and high spatial resolution \citep[e.g., observations of HNC, HCN, \ce{H2CO} and \ce{CH3OH};][]{2014ApJ...792L...2C,2017ApJ...837..177C}.  As of 2017, ALMA consists of 43 \SI{12}{\m} antennas and has unique capabilities to probe the physical and chemical structure of the innermost regions of the coma with great accuracy, providing new impetus to theoretical investigations of the coma.  Spatially-resolved observations allow for study of asymmetric outgassing, acceleration and cooling of the coma gases, and variations in gas kinetic temperature resulting in greatly improved accuracy of the derived molecular abundances.  In addition the release site of a given cometary species, whether the nucleus or an extended source in the coma, can be constrained by measuring the spatial distribution in the inner coma on scales of the order of \SI{1000}{\km} from the nucleus for a compact array configuration.

Beginning in the second half of ALMA observing cycle 5 in  early 2018, the newly installed Band 5 dual polarisations receivers, covering the frequency range \SIrange{157}{212}{\GHz}, will become available creating exciting new observational possibilities.  By providing a sensitive access to rotational transitions from \ce{H2O} and its isotopologues, ALMA has the capability to measure isotopic ratios in \ce{H2O} to test for the thermal and radiative processing history.  These observations will contribute to our understanding of where and how cometary materials originated, as well as offering insight into the physical and chemical conditions of the solar nebula. Based on the upper limits on the \ce{H2O} of \SIlist{<4e25;<7.6e25}{\mols} for MBCs 176P and P/2012 T1 obtained by \herschel{} \citep{devalborro12,orourke13}, the total estimated time for a \SIlist{4e25}{\mols} detection of the \ce{H2O} $3_{13}$--$2_{20}$ emission line at \SI{183.310}{\GHz} with ALMA is \SI{80}{\hour} including calibration overheads.  These observations require excellent weather conditions with precipitable water vapor in the atmosphere \SI{<0.4}{\mm}. Therefore, ALMA does not offer a promising way to detect water in MBCs for sensitivity reasons, unless a new MBC is discovered that is 10 times brighter than the ones observed by \herschel{}.

The next generation of Extremely Large Telescopes (ELTs) are primarily designed to work at red optical and NIR wavelengths. For the 39~m ESO ELT, currently scheduled to be commissioned in 2024, the low- and intermediate-dispersion HARMONI spectrograph will have a wavelength coverage of $0.47-2.45$ $\mu$m, missing the primary fluorescence bands of OH(0-0) and CN(0-0). The 30~m Giant Magellan Telescope, scheduled for commissioning in 2022, will have the low/intermediate resolution spectrograph GMACS and the high resolution spectrograph G-CLEF, both of which will operate at $0.35-0.95~\mu$m and will be sensitive to CN(0-0). The proposed Thirty Meter Telescope would have the Wide Field Optical Spectrometer (WFOS) as a first light instrument, operating at $0.31-1.0~\mu$m. Hence it would miss the primary OH(0-0) band, although it would allow detection of the OH(1-0) band which has a relative brightness of 5\%--10\% depending on the distance and heliocentric velocity of the MBC. 

\subsection{Space telescopes}

With sensitivity in the NIR region, the JWST, a 6.5 m telescope orbiting the Earth-Sun L2 point, may provide our most immediate opportunity to directly detect water in a MBC coma \citep{kelley16}.  The two brightest fluorescence bands of water are the $\nu_3$ band at 2.7~\micron{} and the $\nu_2$ band at 6.3~\micron{}, with fluorescence band $g$-factors near $3.2\times10^{-4}$ and $2.4\times10^{-4}$~s$^{-1}$, respectively, when opacity effects are neglected \citep{Crovisier1983,Bockelee-Morvan2009,Debout2016}.  JWST has spectral sensitivity at both bands: the NIRSpec instrument covers 2.7~\micron{} at spectral resolving powers, $R=\Delta\lambda/\lambda$, of $\sim$100, $\sim$1000, or $\sim$2700; and MIRI covers 6.3~\micron{} with $R\sim$70 and $\sim$3500 \citep{Kendrew2015, Wells2015}.  In general for comets and dark asteroids, the 2--3~\micron{} region has a lower continuum flux than the 6--7~\micron{} region, due to the low albedos of dust and surfaces.  In addition, the sky background is stronger at the longer wavelengths due to zodiacal dust.  Thus, given the similarities in band $g$-factors, the $\nu_3$ band should be easier to observe in a MBC coma.  We ran a JWST exposure time simulation with a NIRSpec fixed-slit (0.4 arcsec wide) and a $V$=19~mag MBC at $r=2.7$~au, producing water with a rate of $10^{25}$ molec.\,s$^{-1}$.  The continuum would be detected with an SNR of 45 in 1~hour.  Assuming the brightest water lines are about one-tenth the total band flux ($\sim1\times10^{-17}$~erg~s$^{-1}$~cm$^{-2}$ in a 0.2 arcsec radius aperture), they may be detected with a peak signal-to-noise ratio of about 3 above the continuum in $10^4$~s of integration time.  This is a challenging observation, but with the right target and production rate, a direct water detection should be possible.

A Far-Infrared (Far-IR) Surveyor mission concept for NASA's 2020 Astrophysics Decadal Survey has been proposed recently \citep{2016SPIE.9904E..0KM}. Originally envisaged in NASA's 2013 Roadmap, this mission intends to build upon previous successful space and airborne infrared astronomy observatories such as ISO, Spitzer, Herschel, SOFIA and JWST.  The mission will have a single large aperture telescope or an interferometer with a large gain in sensitivity of about \numrange{e3}{e4} over the Herschel Space Observatory, and better angular resolution.  The Far-IR Surveyor will be able to detect both water vapour and water ice, using the 43 $\mu$m water-ice band and a large number of water bands, as well as other volatile molecules. In small bodies in the Solar System, the Far-IR Surveyor can be used to detect gas sublimation and accurately measure the water production rates. Direct detection of water in MBCs will be possible with sensitivity to production rates of \SI{e22}{\mols} at a heliocentric distance of 2.5 au. Alternatively, preliminary work continues for a future large UV/optical space telescope, with a 10--15 m diameter. Variously known as HDST or LUVIOR, details are not yet well known, but this will be very sensitive to many gas emission features, and will certainly be useful in studying weakly active comets, but is many years from becoming reality \citep{HDST}.

An alternative approach to building increasingly large space telescopes is to take a small space telescope closer to the MBCs. CASTAway was recently proposed (as an ESA medium class mission) to launch a 50 cm diameter telescope, equipped with a low resolution spectrograph covering 0.6--5 $\mu$m and a CCD camera, into an orbit that loops through the asteroid belt \citep{castaway}. This mission is primarily designed to map the diversity of bodies in the asteroid belt, by performing a spectroscopic survey of more than 10,000 asteroids of all sizes and close flybys of 10 -- 20 diverse objects. Even if a MBC cannot be one of the flyby targets (the relatively small size of the population makes it unlikely that a suitable multi-asteroid tour can include one), the telescope is designed to search for ice or outgassing, in the NIR through spectroscopy or via OH emission in the UV through narrowband imaging. The advantage of a dedicated survey mission, compared with JWST or other major facilities, is that many more potentially ice bodies can be targeted. Placing a telescope in the asteroid belt has the advantage that all sizes of asteroid can be targeted, and also presents unique viewing geometries. For MBCs, such view points can be useful to characterise dust tails, as was demonstrated by imaging P/2010 A2 with the OSIRIS cameras on Rosetta \citep{Snodgrass2010} -- any camera on any mission passing through the asteroid belt can potentially contribute to MBC dust trail characterisation as a target of opportunity.

\subsection{Missions to MBCs}

These various planned and proposed telescope facilities show that, while challenging, the detection of water outgassing from MBCs may be achieved in the near future. 
However, further detailed exploration of the water content of these bodies, and especially more challenging observations such as measurement of isotopic ratios (e.g., D/H in water), will be beyond even the next generation of telescope facilities. Investigating D/H, and the associated constraints on where in the Solar System planetary disc the ice condensed,  will require {\it in situ} measurement. Missions to visit MBCs have been proposed for this purpose, to both ESA and NASA. 

The European proposal Castalia would follow on from ESA's success with Rosetta, and take copies of some of the same key instruments, including the sensitive ROSINA mass spectrometer, to 133P. The mission is described in detail by \citet{castalia}, but in summary this would be a rendezvous mission launched in the late 2020s, to arrive before 133P's 2035 perihelion. Its instrument complement comprises visible, NIR, and thermal IR cameras; mass spectrometers sensitive to neutral and ionised gas; a dust instrument that combines the strengths of the Rosetta GIADA and COSIMA instruments to study flux and composition of grains; a sensitive magnetometer and plasma package; and two ground-penetrating radars. Castalia would perform a detailed characterisation of the nucleus, including quantifying the amount and depth of buried ice via the first sub-surface radar measurements at a small body, and directly `sniff' the escaping gas, measuring its composition at an isotopic level. Combining results from the similar Castalia and Rosetta instruments would allow very direct comparison between a MBC and JFC. As a proposal for the M-class of ESA missions Castalia is necessarily a simpler spacecraft than Rosetta (it doesn't carry a lander, for example), but the low activity and more circular orbit of MBCs helps make the mission `easier'. The proposal is under consideration in September 2017, with the shortlist for phase A study for the ESA M5 call expected to be announced in December 2017.

A similar, but even more bare-essentials approach is taken by the proposed NASA Discovery mission Proteus \citep{proteus}. This mission concentrates on its ability to measure composition and isotopy at very high precision in the low activity MBC environment, with the only other instrument (apart from the sensitive mass spectrometer) being a camera for context and surface characterisation. The core scientific motivation is testing whether or not the water in MBCs has an isotopic match to Earth's oceans. It was proposed to visit 238P in the last NASA Discovery round, and although not selected on that occasion is expected to be reproposed in a future round.

Finally, a proposed mission to an earlier round of ESA M-class missions aimed to not only visit a MBC, but to return a sample of dust from it, capturing this in aerogel during a flyby in the same way the NASA Stardust mission did at comet 81P/Wild~2. Although this mission (Caroline -- \citealt{Caroline}) would not have been sensitive to the volatile component of the MBC, the opportunity to apply sensitive laboratory techniques to grains would certainly be revealing about the origins of the parent body. Currently the {\it in situ} investigation of volatiles is seen as the priority, with many of the Caroline team working on the Castalia proposal instead, but this concept remains interesting to investigate further in the future.


\section{Conclusions}

The repeated activity of MBCs is strong evidence that there is buried ice in numerous small asteroids in the MB, despite surface temperatures being too high for it to be stable. Thermal models suggest that this ice is buried up to $\sim$ 30 m deep, but it is thought that impacts could excavate enough of the insulating layer to allow periods of activity. Outgassing from this activity will produce characteristic spectral features across a wide range of wavelengths, from the UV to radio, but attempts to detect these in MBCs have not yet succeeded. Given the low activity levels expected from MBCs, it appears that this detection is beyond the capabilities of current telescope technology, but may be achievable in the coming years with new facilities, in particular the JWST. The 2.7 $\mu$m and 557 GHz water bands, and possibly UV observations of hydrogen (Ly-$\alpha$) and OH, appear to be the most promising regions of the spectrum to attempt detections. To get more detailed information, such as the isotopic ratios that can answer questions on the relevance of MBCs as a source of Earth's water, or their original formation location in the proto-planetary disc, will require {\it in situ} investigation via a spacecraft visit.

\begin{acknowledgements}
This work is a direct result of support by the International Space Science Institute, Bern, Switzerland through the hosting and provision of financial support for an international team to discuss the science of MBCs. 
CS is supported by the UK STFC as a Rutherford fellow.
AF was supported by UK STFC grant ST/L000709/1.
HHH and MMK were supported by NASA Solar System Observations grant NNX16AD68G. 
MTH is financially supported by David Jewitt. 
MdVB was supported by NASA's Planetary Astronomy Program.
MC is supported by NASA Solar System Observations Grant NNX15AJ81G.
We thank Dave Schleicher for useful discussions. 
We made use of the Planetary Spectrum Generator developed by Geronimo Villanueva at NASA's Goddard Space Flight Center, and thank Geronimo for making this useful service available to the community.
\end{acknowledgements}

\bibliographystyle{spbasic}      
\bibliography{refs02.bib}   

\end{document}